\def\year{2021}\relax
\documentclass[letterpaper]{article} % DO NOT CHANGE THIS
\usepackage{aaai21}  % DO NOT CHANGE THIS
\usepackage{times}  % DO NOT CHANGE THIS
\usepackage{helvet} % DO NOT CHANGE THIS
\usepackage{courier}  % DO NOT CHANGE THIS
\usepackage[hyphens]{url}  % DO NOT CHANGE THIS
\usepackage{graphicx} % DO NOT CHANGE THIS

\usepackage{subfigure}
\usepackage{xcolor}
\usepackage{amsmath,amssymb,amsfonts}
\usepackage{algorithm, algorithmic}
\usepackage{bm}
\usepackage{setspace}
\usepackage{appendix}
\usepackage[switch]{lineno} 

\usepackage{natbib}  % DO NOT CHANGE THIS AND DO NOT ADD ANY OPTIONS TO IT
\usepackage{caption} % DO NOT CHANGE THIS AND DO NOT ADD ANY OPTIONS TO IT
\usepackage{array}
\newcolumntype{P}[1]{>{\centering\arraybackslash}p{#1}}

\usepackage{booktabs}
\usepackage{multirow}

\usepackage{pifont}

\title{Neural Latent Space Model for Dynamic Networks and \\ Temporal Knowledge Graphs}

\author{
    Tony Gracious,\textsuperscript{\rm 1}\thanks{Equal contribution.}
    Shubham Gupta,\textsuperscript{\rm 1}\footnotemark[1]
    Arun Kanthali,\textsuperscript{\rm 1}
    Rui M. Castro,\textsuperscript{\rm 2}
    Ambedkar Dukkipati\textsuperscript{\rm 1} \\
}
\affiliations{
    \textsuperscript{\rm 1}
    Indian Institute of Science, Bangalore - 560012, INDIA. \\

    \textsuperscript{\rm 2}
    Eindhoven University of Technology, 5600 MB Eindhoven, The Netherlands. \\

    [tonygracious, shubhamg, ambedkar]@iisc.ac.in,
    rmcastro@tue.nl 
}

\begin{document}

\maketitle

% =============================================== %

\begin{abstract}
Although static networks have been extensively studied in machine learning, data mining, and AI communities for many decades, the study of dynamic networks has recently taken center stage due to the prominence of social media and its effects on the dynamics of social networks. In this paper, we propose a statistical model for dynamically evolving networks, together with a variational inference approach. Our model, Neural Latent Space Model with Variational Inference, encodes edge dependencies across different time snapshots. It represents nodes via latent vectors and uses interaction matrices to model the presence of edges. These matrices can be used to incorporate multiple relations in heterogeneous networks by having a separate matrix for each of the relations. To capture the temporal dynamics, both node vectors and interaction matrices are allowed to evolve with time. Existing network analysis methods use representation learning techniques for modelling networks. These techniques are different for homogeneous and heterogeneous networks because heterogeneous networks can have multiple types of edges and nodes as opposed to a homogeneous network. Unlike these, we propose a unified model for homogeneous and heterogeneous networks in a variational inference framework. Moreover, the learned node latent vectors and interaction matrices may be interpretable and therefore provide insights on the mechanisms behind network evolution. We experimented with a single step and multi-step link forecasting on real-world networks of homogeneous, bipartite, and heterogeneous nature, and demonstrated that our model significantly outperforms existing models.
\end{abstract}

% =============================================== %

\section{Introduction}
\label{section:introduction}
Network analysis is by no means a new field and consequently, a towering wealth of literature that explores various aspects of network analysis is available \cite{GolderbergEtAl:2010:ASurveyOfStatisticalNetworkModels}. With the advent of deep learning, network analysis methods more recently started focusing on representation learning techniques. These methods learn finite vector representations or embeddings for nodes and edges that can be used for downstream tasks like link prediction \cite{GroverLeskovec:2016:Node2VecScalableFeatureLearningForNetworks}, node classification \cite{SenEtAL:2008:CollectiveClassificationinNetworkData}, community detection \cite{Fortunato:2010:CommunityDetectionInGraphs}, and so on. These techniques are of two kinds. The first kind is for homogeneous networks with only a single type of nodes and edges. The second kind is for heterogeneous networks with multiple types of nodes and edges. These networks can also be viewed as Knowledge Graphs (KG) with different types of nodes as entities and different types of edges as relations between entities. 

Early works on homogeneous network representation learning use random walk models to capture the neighborhood context \cite{PerozziEtAL:2014:DeepWalk:OnlineLearningofSocialRepresentations, GroverLeskovec:2016:Node2VecScalableFeatureLearningForNetworks}. The embeddings are learned so that nodes closer to each other have similar embeddings. In a KG, real-world facts are stored using edges which are represented as a triplet of the form \texttt{(Subject Entity, Relation, Object Entity)}. Here, \texttt{Subject Entity} and \texttt{Object Entity} form the nodes, and \texttt{Relation} is the type of edge. The modelling techniques developed for homogeneous networks are not applicable for KGs as the entities connected via a relation may not be similar. For example, consider the edge \texttt{(Barack Obama, Born in , USA)} which represents a relation \texttt{Born in} between the entities \texttt{Barack Obama} and \texttt{USA}. Here, entities are not similar because \texttt{Barack Obama} is a person and \texttt{USA} is a country. For representation learning in such heterogeneous networks, entities and relations are given a finite representation which is then used as input to a scoring function as in \cite{BordesEtAl:2013:TranslatingEmbeddingsforModelingMulti-relationalData, DettmersEtAL:2018:Convolutional2dknowledgegraphembeddings, ZhigingEtAL:2019:RotatE:KnowledgeGraphEmbeddingbyRelationalRotationinComplexSpace}. The embeddings are learned so that the observed edges in the KG get a higher score as compared to the unobserved edges. 

One aspect that is often ignored by the existing methods is that most real-world networks evolve with time. For example, in social networks, new friendship links are formed or broken with time. Similar observations can be made in KGs as there are relations with temporal properties. For example, \texttt{(Barack Obama, President of, USA)} is valid only between $2009$ and $2017$. KGs which encode such facts are called temporal KGs. Here, the links change with time. Owing to many practical applications, it is important to integrate both dynamic and heterogeneous information while modelling networks.
 
Existing methods for dynamic homogeneous networks have used techniques such as temporal regularization loss \cite{ZhouYangRenWuZhuang:2018:Dynamicnetworkembeddingbymodelingtriadicclosureprocess, GoyalKamraHeLiu:2018:DynGEM:DeepEmbeddingMethodforDynamicGraphs} and RNN based sequential architectures \cite{goyalChhetriCanedo:2020:dyngraph2vec:Capturingnetworkdynamicsusindynamicgraphrepresentationlearning} for modelling the graph evolution. For modelling temporal KGs, previous methods have used temporal point process models  \cite{trivediEtAL:2017:Know-Evolve:DeepTemporalReasoningforDynamicKnowledgeGraphs, trivediEtAL:2019:DyRep:LearningRepresentationsoverDynamicGraphs}, or a RNN based interaction history encoding model \cite{WoojeongHeQuChenZhangSzekelyRe:2019:RecurrentEventNetwork:GlobalStructureInferenceOverTemporalKnowledgeGraph} to predict the future evolution of the networks. To the best of our knowledge, ours is the first method that considers homogeneous and heterogeneous networks in an unified manner.

We propose a statistical model, called Neural Latent Space Model (NLSM), for dynamic networks to address the above-mentioned challenges. If a network has $R$ types of relations, our model uses $R$ sets of interaction matrices, one for each type of relation. For homogeneous networks $R = 1$ and for heterogeneous networks $R > 1$. Our approach uses an unified probabilistic framework which can scale up to the complexities in the network structure. In theory, the parameters of this proposed model could be estimated using training data via Bayesian inference. However, the likelihood structure of the model is complex and non-convex, making such methods computationally infeasible. This motivates a neural network-based variational inference procedure yielding an end-to-end trainable architecture that can be used for efficient and scalable inference.

Our main contributions are as follows. \textbf{(i)} We have proposed a new statistical model for dynamic networks that encodes temporal edge dependencies and can model both homogeneous and heterogeneous networks. \textbf{(ii)} We have provided ample empirical evidence to demonstrate that our model is suitable for link forecasting and it may simultaneously provide important insights into the network evolution mechanics via interpretable embeddings. \textbf{(iii)} In dynamic homogeneous networks, we observed an average performance improvement (over existing state-of-the-art) of $4\%$ in   Micro-AUC metric for single-step link forecasting. Similarly, in dynamic bipartite networks, an average performance improvement of $8.7\%$ in Micro-AUC metric for single-step link forecasting was observed. In dynamic heterogeneous networks, the average improvement is $7.9\%$ in the mean reciprocal rank metric for multi-step link forecasting task.

% =============================================== %

\section{Neural Latent Space Model}
\label{section:model_description}

% =============================================== %

\subsection{Modeling Individual Snapshots} 
\label{section:modeling_individual_snapshots}
In our model, time $t\in\{1,2,\ldots, T\}$ is discrete. The network evolution is therefore described by the corresponding network snapshots at each time-step, specified by binary adjacency matrices $\mathbf{A}_r^{(t)} \in \{0, 1\}^{N \times N}$, where $N$ is the number of nodes in the network, $r \in \{1, 2, \dots, R\}$ is the relation between the nodes, and $t$ denotes the time. We begin by discussing the case of homogeneous networks ($R = 1$). The extension to heterogeneous networks ($R \geq 2$) is then straightforward and we present it in Section \ref{section:modeling_heterogeneous_networks}. To avoid cluttering the notation, we drop the subscript in $\mathbf{A}_r^{(t)}$ when $R = 1$. We further assume that there are no self-loops. Each node is modeled by $K$ attributes whose values lie in the interval $[0, 1]$. These attributes can change over time. The latent vector $\mathbf{z}_n^{(t)} \in [0, 1]^K$ is used to denote the attributes for node $n$ at time $t$.

The interaction between latent vectors of each pair of nodes directly dictates the probability of observing an edge between them. For simplicity, our interaction model encodes only interactions between attributes of the same type, described by \textit{interaction matrices}. For homogeneous networks let $\bm{\Theta}_k^{(t)} \in \mathbb{R}^{2 \times 2}$ be a matrix that encodes the affinity between nodes with respect to attribute $k$ at time $t$. At the time $t$, the node latent vector and interaction matrices fully determine the probability of edges being present. Formally, given $\bm{\Theta}_k^{(t)}$, $k =1,\ldots,K$ and the latent vectors for all nodes $\mathbf{z}_{n}^{(t)}$, $n = 1,\ldots,N$, edges occur independently and the probability of an edge from node $i$ to node $j$ is modeled as:
\begin{equation}
    \label{eq:modified_edge_prob}
    P\left(a_{ij}^{(t)} = 1 | \mathbf{z}_i^{(t)}, \mathbf{z}_j^{(t)}, \{\bm{\Theta}_k^{(t)}\}_{k=1}^K\right) = \sigma\left(\sum_{k=1}^{K} \tilde{\theta}_k^{(t)}(i, j)\right),
\end{equation}
where, $\tilde{\theta}_k^{(t)}(i, j)$ is defined as:
\begin{equation}
    \label{eq:tilde_theta_j_definition_static}
    \tilde{\theta}_k^{(t)}(i, j) = \mathsf{E}_{x \sim B(z_{ik}^{(t)}), y \sim B(z_{jk}^{(t)})}\left[\Theta_k^{(t)}(x, y)\right].
\end{equation}
Here $\sigma(.)$ is the \textit{sigmoid} function, $B(\alpha)$ refers to a Bernoulli distribution with parameter $\alpha$ and $\Theta_k^{(t)}(x, y)$ is the $(x, y)^{th}$ entry of $\Theta_k^{(t)}$ matrix. Note that $x$ and $y$ are independent in \eqref{eq:tilde_theta_j_definition_static}. This formulation allows representation of both homophilic and heterophilic interactions among nodes depending on the structure of the matrices $\Theta_k^{(t)}$. For the case of undirected graphs, the matrices $\Theta_k^{(t)}$ are symmetric.

The interaction model we consider is in the same spirit as the Multiplicative Attribute Graph (MAG) model \cite{KimLeskovec:2012:MultiplicativeAttributeGraphModelOfRealWorldNetworks}. Some other dynamic network models \cite{KimEtAl:2013:NonparametricMultiGroupMembershipModelForDynamicNetworks} use the MAG model directly to represent each static network snapshot, however, in our case, we have a few differences: our node attributes are not restricted to being binary and we have a differentiable expectation operation as given in \eqref{eq:tilde_theta_j_definition_static} instead of the non-differentiable ``selection'' operation given in \cite{KimLeskovec:2012:MultiplicativeAttributeGraphModelOfRealWorldNetworks}. These differences are crucial for one to use a neural network-based variational inference procedure.

% =============================================== %

\subsection{Modeling Network Dynamics}
\label{section:modeling_network_dynamics}
Having described how each network snapshot is generated, it remains to describe how attributes and their interactions evolve over time. To make an analogy with genetics, each attribute type might be seen as a \textit{gene}, and the latent vector corresponds to the \textit{gene expression profile} of a given node. The level of expression of each attribute might change over time - nodes may start exhibiting new attributes and stop exhibiting old ones thereby leading to a change in $\mathbf{z}_n^{(t)}$. At the same time, the role of each attribute in regulating the presence of edges in the network may also change over time leading to a change in $\bm{\Theta}_k^{(t)}$ matrices.

One approach to model the dynamics of a network is to use domain expertise to impose a specific set of assumptions on the process governing the dynamics. However, this limits the class of networks that can be faithfully modeled. Instead, we adopt the strategy of imposing a minimal set of assumptions on the dynamics. This is in the same spirit as in the models used in tracking using stochastic filtering (e.g., Kalman filters and variants) \cite{YilmazJavedShah:2006:ObjectTrackingASurvey}, where dynamics are rather simple and primarily capture the insight that the state of the system cannot change too dramatically over time. The use of simple dynamics together with a powerful function approximator (a neural network) during the inference ensures that a simple yet powerful model can be learned from observed network data.

Let $\bar{\bm{\theta}}_k^{(t)}$ be a vector consisting of the entries of the $\bm{\Theta}_k^{(t)}$ matrix\footnote{For directed graphs the matrix $\bm{\Theta}_k^{(t)}$ can be arbitrary, therefore $\bar{\bm{\theta}}_k^{(t)}$ will have four entries. In the case of undirected graphs the matrices are symmetric, and therefore three entries suffice.}. We model the evolution of interaction matrices as:
\begin{equation}
    \label{eq:interaction_matrices_evolution}
    \bar{\bm{\theta}}_k^{(t)} \sim \mathcal{N}(\bar{\bm{\theta}}_k^{(t - 1)}, s_\theta^2 \mathbf{I}) , \,\, k = 1, ..., K, \,t = 2, ..., T,
\end{equation}
independently over time, where $s_\theta^2 \in \mathbb{R}^+$ is a model hyperparameter and $\mathbf{I}$ denotes the identity matrix. This model captures the intuition that the interaction matrices will likely not change dramatically over time.

Since the entries of the latent vector $\mathbf{z}_n^{(t)}$ are restricted to lie in $[0,1]$, a similar dynamics model as above is not possible. A simple workaround is to re-parameterize the problem by introducing the vectors $\bm{\psi}_n^{(t)} \in \mathbb{R}^K$ such that
\begin{equation}
    \label{eq:relation_between_psi_and_z}
    z_{nk}^{(t)} = \sigma(\psi_{nk}^{(t)}).
\end{equation}
As before, $\sigma(.)$ is the sigmoid function. Now we can have an evolution model similar to (\ref{eq:interaction_matrices_evolution}) on vectors $\bm{\psi}_n^{(t)}$. That is,
\begin{equation}
    \label{eq:node_attributes_evolution}
    \bm{\psi}_n^{(t)} \sim \mathcal{N}(\bm{\psi}_n^{(t - 1)}, s_\psi^2 \mathbf{I}) \,\, n = 1, ..., N, \,t = 2, ..., T,
\end{equation}
where $s_\psi^2$ is a model hyperparameter. This in turn models the evolution of vectors $\mathbf{z}_n^{(t)}$.

Note that (\ref{eq:interaction_matrices_evolution}) and (\ref{eq:node_attributes_evolution}) only imply that the values of variables are unlikely to change very quickly. Other than that, they do not place any strong or network-specific restriction on the dynamics. The hyperparameters $s_\theta^2$ and $s_\psi^2$ control the radius around the current value of the variable within which it is likely to stay in the next timestep.

This approach for modeling dynamics has advantages and disadvantages. The major advantage is flexibility since during inference time, a powerful enough function approximator can learn appropriate network dynamics from the observed data. However, if we regard this proposal as a generative model, then it will lead to unrealistic global behavior. Nonetheless, locally (in time) it will capture the type of dynamics one sees in many networks, and this is enough to ensure good tracking performance. In many real-world cases, a suitable amount of observed data is available but clues about the network dynamics are unavailable. Since the task is to gain meaningful insights from the data, we believe the advantages of this approach outweigh the disadvantages.

Note that \eqref{eq:interaction_matrices_evolution} and \eqref{eq:node_attributes_evolution} are applicable from timestep $2$ onward. We need a way to obtain the initial values for $\bm{\psi}_n^{(1)}$ and $\bar{\bm{\theta}}_k^{(1)}$ for $n = 1, 2, ..., N$ and $k = 1, 2, ..., K$. The initial vectors $\bm{\psi}_n^{(1)}$ and $\bar{\bm{\theta}}_k^{(1)}$ are sampled from a prior distribution. We use the following prior distributions:
\begin{equation}
    \label{eq:theta_prior}
    \bar{\bm{\theta}}_k^{(1)} \sim \mathcal{N}(\mathbf{0}, \sigma_\theta^2\mathbf{I}),
\end{equation}
\begin{equation}
    \label{eq:psi_prior}
    \bm{\psi}_n^{(1)} \sim \mathcal{N}(\mathbf{0}, \sigma_\psi^2\mathbf{I}).
\end{equation}
Here, $\sigma_\theta$ and $\sigma_\psi$ are hyperparameters. In our experiments, we set these hyperparameters to a high value ($\sigma_\theta=\sigma_\psi=10$). This allows the initial embeddings to become flexible enough to represent the first snapshot faithfully. After that, the assumption that the network changes slowly ((\ref{eq:interaction_matrices_evolution}) and (\ref{eq:node_attributes_evolution})) is used to sample the value of random variables $\bm{\psi}_n^{(t)}$ and $\bar{\bm{\theta}}_k^{(t)}$ for $t = 2, 3, ..., T$.

We make the following independence assumptions: given $\bm{\psi}_n^{(t - 1)}$ the vectors $\bm{\psi}_n^{(t)}$ are independent of any quantity indexed by time $t'\leq t-1$. An analogous statement applies to the interaction matrices $\bar{\bm{\theta}}_k^{(t)}$. Finally, given $\bm{\psi}_i^{(t)}$, $\bm{\psi}_j^{(t)}$ and $\bar{\bm{\Theta}}^{(t)} = \{\bar{\bm{\theta}}_k^{(t)}\}_{k=1}^K$, the entries $a_{ij}^{(t)}$ are independent of everything else. The graphical model for NLSM is given in Appendix A in the supplementary material \cite{ThisPaperSupp}. 
Algorithm~\ref{alg:generative_process} outlines the generative process for NLSM.

% =============================================== %

\subsection{Modeling Heterogeneous Networks}
\label{section:modeling_heterogeneous_networks}

In a heterogeneous network, the nodes may have different \textit{types} of relationships between them. A classic example is a knowledge graph where, for instance, \texttt{is president of} and \texttt{lives in} relations have different semantics. The set of nodes, and hence the latent vectors $\mathbf{z}_n^{(t)}$, remains unchanged. To model different types of relations, we use relation specific interaction matrices. Hence, the interaction matrix for $k^{th}$ attribute now depends on the relation $r$, and we denote this by $\bm{\Theta}^{(t)}_{k, r}$, for $k = 1, 2, \dots, K$, and $r = 1, 2, \dots, R$. To compute the edge probabilities under a specified relation $r$, \eqref{eq:modified_edge_prob} only uses interaction matrices of type $r$. Similarly, \eqref{eq:tilde_theta_j_definition_static}, \eqref{eq:interaction_matrices_evolution}, and \eqref{eq:theta_prior} individually apply to each $r$ dependent interaction matrix. Using common latent attributes for nodes captures the required relationships among relations, hence the interaction matrices can evolve independently of each other. In a graph with several types of nodes, because all nodes share the same set of attributes, $K$ must be large enough to accommodate the diverse properties that must be encoded. Naturally, not all elements of the node vectors will be relevant to all node types.

\begin{algorithm}[tb]
\caption{Generative process for NLSM}
\label{alg:generative_process}
\begin{algorithmic}
    \STATE {\bfseries Input:} ${N}$: Number of nodes, \\
    \hspace{11mm}${K}$: Latent vector dimension, \\
    \hspace{11mm}${T}$: Number of timesteps, \\
    \hspace{11mm}$s_\theta^2$: Hyperparameter used in \eqref{eq:interaction_matrices_evolution}, \\
    \hspace{11mm}$s_\psi^2$: Hyperparameter used in \eqref{eq:node_attributes_evolution}, \\
    \hspace{11mm}$\sigma_\theta^2$: Hyperparameter used in \eqref{eq:theta_prior}, and \\
    \hspace{11mm}$\sigma_\psi^2$: Hyperparameter used in \eqref{eq:psi_prior}
    \STATE Sample $\bm{\psi}_n^{(1)}$ using (\ref{eq:psi_prior}) for $n = 1, 2, ..., N$
    \STATE Sample $\bar{\bm{\theta}}_k^{(1)}$ using (\ref{eq:theta_prior}) for $k = 1, 2, ..., K$
    \FOR{$t=1$ {\bfseries to} ${T-1}$}
        \STATE Compute $\mathbf{z}_n^{(t)}$ by using $\bm{\psi}_n^{(t)}$ in (\ref{eq:relation_between_psi_and_z}) for $n = 1, 2, ..., N$
        \STATE Sample $a_{ij}^{(t)}$ using (\ref{eq:modified_edge_prob}) for $i, j = 1, 2, ..., N$, $i \neq j$
        \STATE Sample $\bm{\psi}_n^{(t + 1)}$ using (\ref{eq:node_attributes_evolution}) for $n = 1, 2, ..., N$
        \STATE Sample $\bar{\bm{\theta}}_k^{(t + 1)}$ using (\ref{eq:interaction_matrices_evolution}) for $k = 1, 2, ..., K$
    \ENDFOR
    \STATE Compute $\mathbf{z}_n^{(T)}$ by using $\bm{\psi}_n^{(T)}$ in (\ref{eq:relation_between_psi_and_z}) for $n = 1, 2, ..., N$
    \STATE Sample $a_{ij}^{(T)}$ using (\ref{eq:modified_edge_prob}) for $i, j = 1, 2, ..., N$, $i \neq j$      
    \STATE {\bfseries Return:} $[\mathbf{A}^{(1)}, \mathbf{A}^{(2)}, ..., \mathbf{A}^{(T)}]$
\end{algorithmic}
\end{algorithm}

% =============================================== %

\section{Inference in NLSM}
\label{section:inference}
As before, to maintain readability, we describe the inference procedure for the case of $R = 1$. The general case of $R \geq 1$ trivially follows along the same lines. In practice, an observed sequence of network snapshots $\mathbf{A} = [\mathbf{A}^{(1)}, \mathbf{A}^{(2)}, ..., \mathbf{A}^{(T)}]$ is available, and the main inference task is to estimate the values of the underlying latent random variables. Performing exact inference in NLSM is intractable because the computation of marginalized log probability of observed data results in integrals that are hard to evaluate. Thus, we adopt approximate inference techniques.

Our goal is to compute an approximation to the true posterior distribution $P(\{\bm{\Psi}^{(t)}, \bm{\Theta}^{(t)}\}_{t=1}^T | \{\mathbf{A}^{(t)}\}_{t=1}^T)$. Note that in our approach $K$, $s_\theta$, $s_\psi$, $\sigma_\theta$ and $\sigma_\psi$ are hyperparameters that are simply set by the user. We pose the inference problem as an optimization problem by using Variational Inference \cite{BleiEtAl:2017:VariationalInferenceAReviewForStatisticians} and parameterize the approximating distribution by a neural network. There are several benefits like efficiency and scalability \cite{BleiEtAl:2017:VariationalInferenceAReviewForStatisticians} associated with the use of variational inference. Also, coupled with powerful neural networks, variational inference can model complicated distributions \cite{KingmaEtAl:2013:AutoEncodingVariationalBayes}.

The main idea behind variational inference is to approximate the posterior distribution by a suitable surrogate. Consider a general latent variable model with the set of all observed random variables $\mathbf{X}$ and the set of all latent random variables $\mathbf{H}$. The (intractable) posterior distribution $P(\mathbf{H} | \mathbf{X})$ is approximated by using a parameterized distribution $Q_{\bm{\Phi}}(\mathbf{H})$ where $\bm{\Phi}$ is the set of all the parameters of $Q$. One would like the distribution $Q$ to be as \textit{close} to the distribution $P(\mathbf{H} | \mathbf{X})$ as possible. In general, Kullback-Leibler (KL) divergence is used as a measure of similarity between the two distributions. The goal of variational inference is to find the parameters $\bm{\Phi}$ for which $\text{KL}(Q_{\bm{\Phi}}(\mathbf{H}) || P(\mathbf{H} | \mathbf{X}))$ is minimized. However, this optimization objective is intractable since one cannot efficiently compute $P(\mathbf{H} | \mathbf{X})$. Nevertheless one can show that maximizing the \textit{Evidence Lower Bound Objective (ELBO)} given by \begin{equation}
    \label{eq:elbo}
    \text{ELBO}(\bm{\Phi}) = \mathsf{E}_Q[\log P(\mathbf{X}, \mathbf{H}) - \log Q_{\bm{\Phi}}(\mathbf{H})],
\end{equation}
is equivalent to minimizing the KL criterion \cite{BleiEtAl:2017:VariationalInferenceAReviewForStatisticians} (see Appendix B in the supplementary material \cite{ThisPaperSupp} for proof). For most models, the ELBO can be efficiently computed or approximated by imposing a suitable set of assumptions on $Q$ as described later. We parameterize the distribution $Q$ by a neural network and hence $\bm{\Phi}$ represents the set of parameters of that neural network in our setting.

% =============================================== %

\subsection{Approximating ELBO}
\label{section:approximating_elbo}
The latent variables in our model correspond to the elements of $\bm{\Theta}^{(t)}$ and $\bm{\Psi}^{(t)}$ for $t = 1, 2, ..., T$. The observed variables are $\mathbf{A}^{(1)}, ..., \mathbf{A}^{(T)}$. The parameter vector $\bm\Phi$ consists of the weights of the neural network. Following (\ref{eq:elbo}), we get:
\begin{align}
    \label{eq:elbo_general}
    \text{ELBO}(\bm{\Phi}) = \mathsf{E}_Q\big[&\log P\big(\{\mathbf{A}^{(t)}\}_{t=1}^T, \{\bm{\Psi}^{(t)}, \bm{\Theta}^{(t)}\}_{t=1}^T\big) \nonumber \\ 
    &- \log Q_{\bm{\Phi}}\big(\{\bm{\Psi}^{(t)}, \bm{\Theta}^{(t)}\}_{t=1}^T\big)\big].
\end{align}
Using the independence assumptions stated in Section \ref{section:model_description}, one can write:
\begin{align}
    \label{eq:joint_prob}
    \log &P\big(\{\mathbf{A}^{(t)}\}_{t=1}^T, \{\bm{\Psi}^{(t)}, \bm{\Theta}^{(t)}\}_{t=1}^T\big) = \nonumber \\
    \sum_{n=1}^N &\log P(\bm{\psi}_n^{(1)}) + \sum_{k=1}^K \log P(\bar{\bm{\theta}}_k^{(1)}) + \nonumber \\
    \sum_{t=2}^T &\Big(\sum_{n=1}^N \log P(\bm{\psi}_n^{(t)}|\bm{\psi}_n^{(t-1)}) + \sum_{k=1}^K \log P(\bar{\bm{\theta}}_k^{(t)}|\bar{\bm{\theta}}_k^{(t-1)}) \Big) \nonumber \\
    + &\sum_{t=1}^T \sum_{i \neq j} \log P(a_{ij}^{(t)} | \bm{\psi}_i^{(t)}, \bm{\psi}_j^{(t)}, \bm{\Theta}^{(t)}).
\end{align}
The right hand side of (\ref{eq:joint_prob}) can be computed using (\ref{eq:modified_edge_prob}), (\ref{eq:interaction_matrices_evolution}), (\ref{eq:relation_between_psi_and_z}),  (\ref{eq:node_attributes_evolution}), (\ref{eq:theta_prior}) and (\ref{eq:psi_prior}). Following the standard practice \cite{BleiEtAl:2017:VariationalInferenceAReviewForStatisticians}, we also assume that $Q_{\bm{\Phi}}(.)$ belongs to a mean field family of distributions, i.e. all the variables are independent under $Q$:
\begin{align}
    \label{eq:q_mean_field}
    Q_{\bm{\phi}}\big(\{\bm{\Psi}^{(t)}, \bm{\Theta}^{(t)}\}_{t=1}^T\big) = &\Big(\prod_{t=1}^{T}\prod_{n=1}^N q_{\bm{\psi}_n}^{(t)}(\bm{\psi}_n^{(t)})\Big) \nonumber \\
    &\Big(\prod_{t=1}^{T}\prod_{k=1}^K q_{\bar{\bm{\theta}}_k}^{(t)}(\bar{\bm{\theta}}_k^{(t)})\Big).
\end{align}

We model the distributions $q_{\bm{\psi}_n}^{(t)}$ and $q_{\bar{\bm{\theta}}_k}^{(t)}$ using a Gaussian distribution as given in (\ref{eq:q_psi_gaussian}) and (\ref{eq:q_theta_gaussian}) (with some abuse of notation, $\mathcal{N}$ denotes the density of a normal distribution\footnote{Define $\mathcal{N}(\mathbf{x} | \bm{\mu}, \bm{\Sigma}) = \frac{1}{(2\pi|\bm{\Sigma}|)^{|\bm{\mu}|/2}} \exp{(-\frac{1}{2} d(\mathbf{x}, \bm{\mu}))}$, where $d(\mathbf{x}, \bm{\mu}) = (\mathbf{x} - \bm{\mu})^{\intercal} \Sigma^{-1} (\mathbf{x} - \bm{\mu})$}).
\begin{equation}
    \label{eq:q_psi_gaussian}
    q_{\bm{\psi}_n}^{(t)}(\bm{\psi}_n^{(t)}) = \mathcal{N}(\bm{\psi}_n^{(t)} | \mathbf{m}_{\bm{\psi}_n}^{(t)}, (\bm{\sigma}_{\bm{\psi}_n}^{(t)})^2 \mathbf{I}), \text{and}
\end{equation}
\begin{equation}
    \label{eq:q_theta_gaussian}
    q_{\bar{\bm{\theta}}_k}^{(t)}(\bar{\bm{\theta}}_k^{(t)}) = \mathcal{N}(\bar{\bm{\theta}}_k^{(t)} | \mathbf{m}_{\bar{\bm{\theta}}_k}^{(t)}, (\bm{\sigma}_{\bar{\bm{\theta}}_k}^{(t)})^2 \mathbf{I}).
\end{equation}
Here $(\bm{\sigma}_{x}^{(t)})^2 \mathbf{I} = \text{diag}\big((\sigma_{x}^{(t)})^2_1, ..., (\sigma_{x}^{(t)})^2_{|x|}\big)$. We wish to learn the mean and covariance parameters of Gaussian distributions in \eqref{eq:q_psi_gaussian} and \eqref{eq:q_theta_gaussian} (these are called \textit{variational parameters}). There are two possible approaches for doing this: \textbf{(i)} $\text{ELBO}(\bm{\Phi})$ can be directly optimized as a function of variational parameters or \textbf{(ii)} One can model the variational parameters as outputs of some other parametric function (like a neural network) and then optimize the parameters of that parametric function. The second approach can be viewed as a form of regularization where the space in which variational parameters can lie is constrained to the range of the parametric function in use. We adopt the latter approach, and obtain the variational parameters as outputs of neural networks. We use $\bm{\Phi}$ to denote the set of neural network parameters. Thus $q_{\bm{\psi}_n}^{(t)}(\bm{\psi}_n^{(t)}) \equiv q_{\bm{\psi}_n}^{(t)}(\bm{\psi}_n^{(t)}; \bm{\Phi})$, but we do not explicitly mention the dependence on $\bm{\Phi}$ in general to avoid further notational clutter. The $\text{ELBO}(\bm{\Phi})$ can now be computed by using (\ref{eq:joint_prob}) and (\ref{eq:q_mean_field}) in (\ref{eq:elbo_general}). Integration of the term involving $\log P(a_{ij}^{(t)} | \bm{\psi}_i^{(t)}, \bm{\psi}_j^{(t)}, \bm{\Theta}^{(t)})$ is hard, so for this term we use Monte-Carlo estimation. In all our experiments we use only one sample to get an approximation to (\ref{eq:elbo_general}) as also proposed in \cite{KingmaEtAl:2013:AutoEncodingVariationalBayes}. Additionally, we observed in our experiments that for $t=1$, using $\mathbf{m}_{\bm{\psi}_n}^{(1)}$ and $\bar{\bm{\theta}}_k^{(1)}$ directly as a sample for Monte-Carlo estimation improves the performance for link forecasting and hence we do this in all our experiments.

% =============================================== %

\subsection{Network Architecture}
\label{section:network_architecture}
We use a neural network to parameterize the distributions in \eqref{eq:q_psi_gaussian} and \eqref{eq:q_theta_gaussian}. Our network consists of four GRUs \cite{ChoEtAl:2014:OnThePropertiesOfNeuralMachineTranslationEncoderDecoredApproaches}, one each for the mean and covariance parameters ($\mathbf{m}_{\bm{\psi}}$, $\bm{\sigma}_{\bm{\psi}}$, $\mathbf{m}_{\bar{\bm{\theta}}}$ and $\bm{\sigma}_{\bar{\bm{\theta}}}$). We refer to these GRUs as $\mathcal{G}_{\mathbf{m}}^{\bm{\psi}}$, $\mathcal{G}_{\bm{\sigma}}^{\bm{\psi}}$, $\mathcal{G}_{\mathbf{m}}^{\bar{\bm{\theta}}}$ and $\mathcal{G}_{\bm{\sigma}}^{\bar{\bm{\theta}}}$ respectively. These GRUs interact with each other only during the computation of $\text{ELBO}(\bm{\Phi})$ since their outputs are used to compute \eqref{eq:elbo_general}. See Appendix A in the supplementary material for a visual description.

For brevity of exposition, we will only describe the inputs and outputs for $\mathcal{G}_{\mathbf{m}}^{\bm{\psi}}$. Similar ideas have been employed for other GRUs. For $t=1, 2, ..., T - 1$, $\mathcal{G}_{\mathbf{m}}^{\bm{\psi}}$ generates $\mathbf{m}_{{\bm{\psi}}_n}^{(t + 1)}$ at timestep $t$ for all nodes in the current batch as output. In GRUs, the output of current timestep is used as the input hidden state for the next timestep, thus the input hidden state at timestep $t$ corresponds to $\mathbf{m}_{{\bm{\psi}}_n}^{(t)}$. To be consistent with this, the initial hidden state of $\mathcal{G}_{\mathbf{m}}^{\bm{\psi}}$ is set to $\mathbf{m}_{{\bm{\psi}}_n}^{(1)}$. This means that the initial hidden state for $\mathcal{G}_{\mathbf{m}}^{\bm{\psi}}$ is a learnable vector.

In all our experiments, we use an all $0$'s input vector for $\mathcal{G}_{\mathbf{m}}^{\bm{\psi}}$ at each timestep. If observable features of nodes (that may be dynamic themselves) are available, one can instead use these features as input. For  $\mathcal{G}_{\bm{\sigma}}^{\bm{\psi}}$ and $\mathcal{G}_{\bm{\sigma}}^{\bar{\bm{\theta}}}$, instead of computing the variance terms, which are constrained to be positive, we compute log of variance (this is again standard practice \cite{KingmaEtAl:2013:AutoEncodingVariationalBayes}).

Once the mean and covariance parameters are available, we use the reparameterisation trick \cite{KingmaEtAl:2013:AutoEncodingVariationalBayes} to sample $\bm{\psi}_n^{(t)}$ and $\bar{\bm{\theta}}_k^{(t)}$ using (\ref{eq:q_psi_gaussian}) and (\ref{eq:q_theta_gaussian}) which are then used to approximate $\text{ELBO}(\bm{\Phi})$ using (\ref{eq:elbo_general}) as described in Section \ref{section:approximating_elbo}. The training objective is to maximize $\text{ELBO}(\bm{\Phi})$. The beauty of our model is that $\text{ELBO}(\bm{\Phi})$ is differentiable with respect to $\bm{\Phi}$ and gradients can be easily computed by back-propagation. This means that one can optimize this function using a gradient-based method and therefore capitalize on the powerful optimization methods used for training neural networks. Furthermore, since ELBO uses only pairwise interactions among nodes, we can operate in a batch setting where only a subset of all nodes and the interactions within this subset are considered. This allows us to scale up to rather large networks by training our model on random batches of nodes and their sub-graphs. 

One additional benefit of using a neural network as opposed to learning the variational parameters directly is that the neural network can capture the temporal patterns in the data that can not be captured by the variational parameters on their own, as the unrestricted dynamics model is extremely flexible and can cope with a rather drastic evolution of attributes and interaction matrices. Since the neural network is being trained to predict the parameters for time $t$ given the history up to time $t-1$, it is being encouraged to look for temporal patterns in the data.

In all our experiments we use the well known Adam optimizer \cite{KingmaBa:2014:AdamAMethodforStochasticOptimization} with a learning rate of $0.01$ to train the inference network. A separate inference network is trained for all time steps (in other words, to make predictions for time $t$ we train the inference network with all the observations up to time $t-1$). Note that all networks have exactly the same number of parameters. While training, the parameters of the neural network that is used to make predictions at time $t$ are initialized with the parameters of trained network for time $t - 1$.

% =============================================== %

\begin{table}
\begin{center}
\begin{small}
\begin{sc}
\begin{tabular}{|c|c|c|c|c|}
\hline
\textbf{Dataset} & \textbf{\#Nodes} & \textbf{\#Links} & \textbf{\#Steps} & \textbf{\#Rel} \\
\hline
\textbf{Enron} & 143 & 2,347 & 16  & 1 \\
\textbf{UCI} & 1,809 &  16,822 &  13 & 1\\
\textbf{Yelp} & 6,569 & 95,361 & 16 & 1 \\
\textbf{ML-10M} & 20,537 & 43,760 & 13 & 1 \\
\textbf{YAGO} & 10,623 & 201089 & 186 & 10 \\
\textbf{WIKI} &  12,554 & 669,934 & 232  & 24 \\
\hline
\end{tabular}
\end{sc}
\end{small}
\end{center}
\caption{Dataset Description. Here, \textbf{\#Nodes} is the number of nodes $N$, \textbf{\#Links} is number of edges, \textbf{\#Steps} is the number of snapshots $T$, and \textbf{\#Rel} is the number of relations types $R$.}
\label{table:dataset_description}
\end{table}

\section{Experiments}

% =============================================== %

\subsection{Dataset Description}
\label{section:dataset_description}
We use the UCI \cite{uci2009}, Enron  \cite{Klimt2004IntroducingTE}, Yelp\footnote{https://www.yelp.com/dataset/challenge}, ML-10M \cite{HarperEtAL:2015:TheMovieLensDatasets:HistoryandContext}, WIKI \cite{wiki2018}, and YAGO \cite{Mahdisoltani2014YAGO3AK} datasets in our experiments. Table \ref{table:dataset_description} summarizes the datasets and additional information is provided in the supplementary material.

\begin{table*}
\begin{center}
    \setlength{\tabcolsep}{4pt}
    \begin{tabular}{lcccccccc}
        \textbf{Datasets} & \textbf{AUC}  & \textbf{node2vec}$^a$ & \textbf{DynamiceTraid}$^b$ & \textbf{DynGEM}$^c$ & \textbf{DynAERNN}$^d$ & \textbf{DySAT}$^e$ & \textbf{BAS} & \textbf{NLSM} \\
        \toprule
        \multirow{2}{*}{\textbf{Enron}} & Micro & 83.72 $\pm$ 0.7 & 80.26 $\pm$ 0.8 & 67.83 $\pm$ 0.6 & 72.02 $\pm$ 0.7 & 85.71 $\pm$ 0.3 & 76.88 $\pm$ 0.5 & \textbf{87.05 $\pm$ 0.3} \\
        & Macro & 83.05 $\pm$ 1.2  & 78.98 $\pm$ 0.9  & 69.72 $\pm$ 1.3  & 72.01 $\pm$ 0.7 & \textbf{86.60 $\pm$ 0.2} & 77.62 $\pm$ 0.5 & 86.24 $\pm$ 0.4 \\
        \hline
        \multirow{2}{*}{\textbf{UCI}} & Micro & 79.99 $\pm$ 0.4 & 77.59 $\pm$ 0.6 & 77.49 $\pm$ 0.4 & 79.95 $\pm$ 0.4 & 81.03 $\pm$ 0.2 & 78.79 $\pm$ 0.5 & \textbf{86.24 $\pm$ 0.4} \\
        & Macro & 80.49 $\pm$ 0.6 & 80.28 $\pm$ 0.5 & 79.82 $\pm$ 0.5 & 83.52 $\pm$ 0.4 & 85.81 $\pm$ 0.1 & 83.84 $\pm$ 0.4 & \textbf{88.9 $\pm$ 0.3} \\
        \hline
        \multirow{2}{*}{\textbf{Yelp} } & Micro & 67.86 $\pm$ 0.2 & 63.53 $\pm$ 0.3 & 66.02 $\pm$ 0.2 & 69.54 $\pm$ 0.2 & 70.15 $\pm$ 0.1 & 70.21 $\pm$ 0.1 & \textbf{81.38 $\pm$  0.2} \\
        & Macro & 65.34 $\pm$ 0.2 & 62.69 $\pm$ 0.3 & 65.94 $\pm$ 0.2 & 68.91 $\pm$ 0.2 & 69.87 $\pm$ 0.1 & 69.40 $\pm$ 0.1 & \textbf{80.12 $\pm$ 0.3} \\
        \hline
        \multirow{2}{*}{\textbf{ML-10M}}  & Micro & 87.74 $\pm$ 0.2 & 88.71 $\pm$ 0.2 & 73.69 $\pm$ 1.2 & 87.73 $\pm$ 0.2 & 90.82 $\pm$ 0.3 & 84.10 $\pm$ 0.4 & \textbf{92.21 $\pm$ 0.4} \\
        & Macro & 87.52 $\pm$ 0.3 & 88.43 $\pm$ 0.1 & 85.96 $\pm$ 0.3 & 89.47 $\pm$ 0.1 & \textbf{93.68 $\pm$ 0.1} & 84.32 $\pm$ 0.3 & 92.39 $\pm$ 0.3 \\
        \bottomrule
    \end{tabular}
\end{center}
\caption{Single-step link forecasting results. Micro and Macro AUC in \% averaged over 10 runs with standard deviation. NLSM beats previous state-of-the-art results in almost all settings. Reported performance scores for other methods were taken from \cite{SankarWuGouZhangYang:2020:DySAT:DeepNeuralRepresentationLearningonDynamicGraphsviaSelf-AttentionNetworks}. \textbf{Citations:}$^a$\cite{GroverLeskovec:2016:Node2VecScalableFeatureLearningForNetworks}, $^b$\cite{ZhouYangRenWuZhuang:2018:Dynamicnetworkembeddingbymodelingtriadicclosureprocess}, $^c$\cite{GoyalKamraHeLiu:2018:DynGEM:DeepEmbeddingMethodforDynamicGraphs}, $^d$\cite{goyalChhetriCanedo:2020:dyngraph2vec:Capturingnetworkdynamicsusindynamicgraphrepresentationlearning}, $^e$\cite{SankarWuGouZhangYang:2020:DySAT:DeepNeuralRepresentationLearningonDynamicGraphsviaSelf-AttentionNetworks}
} 
\label{tab:linkpredictiondynamicnetworks}
\end{table*}

% =============================================== %

\subsection{Link Forecasting}
\label{section:link_prediction}
We consider two settings for link forecasting: single-step and multi-step link forecasting. In single-step link forecasting, we are given a dynamic network up to time $t$ as a sequence of snapshots $[\mathbf{A}^{(1)}, \mathbf{A}^{(2)}, \ldots , \mathbf{A}^{(t)}]$ and the task is to predict $\mathbf{A}^{(t + 1)}$. In multi-step link forecasting, the aim is to predict the next $k$ snapshots $\mathbf{A}^{(t + 1)}, \mathbf{A}^{(t+2)}, \dots, \mathbf{A}^{(t+k)}$. 

For evaluating link forecasting in communication and rating networks, we use the well known \textit{Area Under Curve}-score (AUC-score). See Appendix D in the supplementary material for details about evaluation metrics. We compare our performance against existing approaches for both static as well as dynamic network representation learning. While training and evaluating, links formed with nodes that are not observed in the training time-steps are removed.

For Temporal KGs, we use a windowed approach for training the model to reduce the computational requirement as these datasets have a large number of snapshots. We use a moving window of size $m$ having snapshots $\mathbf{A}^{(t -m+1)}, \mathbf{A}^{(t-m+2)}, \dots , \mathbf{A}^{(t)}$ and predict future snapshots. This is done for all $t=2, \dots, T$. The window is shifted by one step if $t > m$. While shifting, we keep the parameters of GRUs fixed but change the initial embeddings of nodes and interaction matrices by initializing them with one step evolved embeddings from their respective GRUs.

We evaluate Temporal KGs in the multi-step link forecasting setting and use well known metrics like Mean Reciprocal Rank (\textbf{MRR}), \textbf{Hits@3}, and \textbf{Hits@10}. We use both raw and filtered versions of these metrics as explained in \cite{BordesEtAl:2013:TranslatingEmbeddingsforModelingMulti-relationalData} to compare our model with existing methods. See Appendix D in the supplementary material for more details about the evaluation metrics.

In all our experiments, we fix the value of $K$ to $64$. This value allows significant flexibility in the model while maintaining computational tractability. Larger values can also be used without significantly affecting the results. Similarly, we chose $s_\theta = s_\psi = 0.1$ and $\sigma_\theta = \sigma_\psi = 10$ for all experiments. Our model is rather robust to the choice of hyperparameters and dataset specific tuning is, in general, not required as exemplified by the fact that we reuse the same values of hyperparameters across all of our experiments.

% =============================================== %

\subsection{Results}
\paragraph{Single Relation Link Forecasting: }
In this setting, we train the model using snapshots until time $t$ and forecast the links occurring at time $t+1$ for each $t=1, \dots, T-1$. We initialize the model for forecasting the links at time $t+1$ with the model trained at time $t$, and then retrain it. We use Micro-AUC and Macro-AUC metrics for evaluating the performance. Micro-AUC is calculated considering all the links across the time-steps and Macro-AUC is the average of AUC at each time-step. These experiments use communication and rating networks and the results are reported in Table ~\ref{tab:linkpredictiondynamicnetworks}. We can see that our model NLSM outperforms all other approaches based on the Micro-AUC scores. We can also observe that our model performs much better as compared to the static network embedding method node2vec. This is because static models do not consider temporal dynamics.

To demonstrate the utility of having GRUs, we created a baseline (BAS). BAS directly approximates $\text{ELBO}(\bm{\Phi})$ as a function of variational parameters in \eqref{eq:q_psi_gaussian} and \eqref{eq:q_theta_gaussian}. This model does not use a GRU for posterior inference. Instead, it learns embeddings for all nodes independently at each time-step. Then, for predicting the future links, it uses the latest embedding of the nodes. Note that BAS performs poorly across datasets. This shows that the regularization provided by the GRUs allows us to better capture the temporal patterns and improve the quality of inference.

\begin{table*}
\begin{center}
\setlength{\tabcolsep}{4.5pt}
\begin{tabular}{lllllllllllll}
    \textbf{Method} & \multicolumn{3}{c}{\textbf{WIKI - filtered}} & \multicolumn{3}{c}{\textbf{WIKI - raw}} & \multicolumn{3}{c}{\textbf{YAGO - filtered}} & \multicolumn{3}{c}{\textbf{YAGO - raw}} \\
    \toprule
    & \textbf{MRR} & \textbf{H@3} & \textbf{H@10} & \textbf{MRR} & \textbf{H@3} & \textbf{H@10} & \textbf{MRR} & \textbf{H@3}   & \textbf{H@10} & \textbf{MRR} & \textbf{H@3} & \textbf{H@1O}  \\
    \toprule
    Know-Evolve+MLP$^a$ & 12.64 & 14.33 &  21.57 & 10.54 & 13.08 & 20.21 & 6.19 & 6.59 & 11.48 &  5.23 & 5.63 & 10.23 \\
    DyRep+MLP$^b$ & 11.60 & 12.74 & 21.65 & 10.41 & 12.06 & 20.93 & 5.87 & 6.54 & 11.98 & 4.98 & 5.54 & 10.19 \\
    RE-NET$^c$ & 53.57 & 54.10 & 55.72 & 32.44 & 35.42 & 43.16 & 66.80 & 67.23 & 69.77 & 48.60 & 54.20 & 63.59 \\
    \midrule
    \textbf{NLSM} & \textbf{56.70} & \textbf{57.80} & \textbf{61.10} & \textbf{35.25} & \textbf{38.60} & \textbf{47.55} & \textbf{69.40} & \textbf{71.25} & \textbf{73.90} & \textbf{52.50} & \textbf{59.20} & \textbf{68.40} \\
    \bottomrule 
\end{tabular}
\end{center}
\caption{Results on Temporal KGs. Reported scores are in \% averaged over 5 runs. Our model performs better than existing approaches. The performance scores for other approaches have been taken from \cite{WoojeongHeQuChenZhangSzekelyRe:2019:RecurrentEventNetwork:GlobalStructureInferenceOverTemporalKnowledgeGraph}. \textbf{Citations: } $^{a}$\cite{trivediEtAL:2017:Know-Evolve:DeepTemporalReasoningforDynamicKnowledgeGraphs}, $^{b}$ \cite{trivediEtAL:2019:DyRep:LearningRepresentationsoverDynamicGraphs}, $^{c}$\cite{WoojeongHeQuChenZhangSzekelyRe:2019:RecurrentEventNetwork:GlobalStructureInferenceOverTemporalKnowledgeGraph} }
\label{tab:linkpredictionTKG}
\end{table*}

\paragraph{Multi-Relational Link Forecasting: }
For multi-relational link forecasting, we use the temporal KGs YAGO and WIKI. We use the heterogeneous variant of our model NLSM as explained in \ref{section:modeling_heterogeneous_networks}. As before, we use a windowed training approach and with window size $m=5$ as in \cite{WoojeongHeQuChenZhangSzekelyRe:2019:RecurrentEventNetwork:GlobalStructureInferenceOverTemporalKnowledgeGraph}. Testing is done by inferring the future embeddings via the node embedding GRUs. We followed the train-test split used by existing works \cite{WoojeongHeQuChenZhangSzekelyRe:2019:RecurrentEventNetwork:GlobalStructureInferenceOverTemporalKnowledgeGraph}. For YAGO we test on the last six time steps and for WIKI we test on the last $10$ time steps. The performance is reported in Table ~\ref{tab:linkpredictionTKG}. It can be observed that our model performs much better than the previous state-of-the-art model RE-NET \cite{WoojeongHeQuChenZhangSzekelyRe:2019:RecurrentEventNetwork:GlobalStructureInferenceOverTemporalKnowledgeGraph} in all the settings.

See Appendix E in the supplementary material \cite{ThisPaperSupp} for additional details regarding our experimental setup. Appendix F presents additional results on smaller datasets and explores the effect of changing $K$. Appendix G presents a qualitative case study to demonstrate that learned embeddings may also offer interpretable insights about the network dynamics.

% =============================================== %

\section{Related Work}
\label{section:related_work}

% =============================================== %

\paragraph{Statistical Network Analysis: }
One of the first successful statistical model for dynamic networks was proposed in \cite{XingEtAl:2010:AStateSpaceMixedMembershipBlockmodelForDynamicNetworkTomography}. It is an extension of the well known Mixed Membership Stochastic Blockmodel \cite{Airoldi:2008:MixedMembershipStochasticBlockmodels} with the additional assumption that parameters evolve via a Gaussian random walk. Since then, multiple researchers have proposed extensions of static network models like Stochastic Blockmodel \cite{HollandEtAl:1983:StochasticBlockmodelsFirstSteps} to the case of dynamic networks \cite{YangEtAl:2011:DetectingCommunitiesAndTheirEvolutionsInDynamicSocialNetworksABayesianapproach,Xu:2014:DynamicStochasticBlockmodelsForTimeEvolvingSocialNetworks,Xu:2015:StochasticBlockTransitionModelsForDynamicNetworks}.

Another class of models extend the general latent space model for static networks to the dynamic network setting \cite{SarkarEtAl:2005:DynamicSocialNetworkAnalysisUsingLatentSpaceModels,FouldsEtAl:2011:ADynamicRelationalInfiniteFeatureModelForLongitudinalSocialNetworks,HeaukulaniEtAl:2013:DynamicProbabilisticModelsForLatentFeaturePropagationInSocialNetworks,KimEtAl:2013:NonparametricMultiGroupMembershipModelForDynamicNetworks,SwellEtAl:2015:LatentSpaceModelsForDynamicNetworks,SwellEtAl:2016:LatentSpaceModelsForDynamicNetworksWithWeightedEdges,GuptaEtAl:2018:EvolvingLatentSpaceModelForDynamicNetworks}. Our proposed model also falls under this category. The basic idea behind such models is to represent each node by an embedding (which may change with time) and model the probability of an edge as a function of the embeddings of the two endpoints. All of these approaches (except \cite{GuptaEtAl:2018:EvolvingLatentSpaceModelForDynamicNetworks}) use an MCMC based inference procedure that does not directly support neural network based inference. However, unlike these previous approaches, in our model the role of each attribute in latent vector can also change. This is a rather distinctive feature of NLSM, allowing us to capture both local dynamics (the evolution of attributes in node latent vector) and global dynamics (the evolving role of attributes). In addition to that, our model for static network snapshots is fully differentiable which allows us to use a neural network based variational inference procedure as opposed to most existing methods that use MCMC based inference. 

% =============================================== %

\paragraph{Dynamic Networks: }
For incorporating temporal properties into embedding learning, methods were devised to model the evolution of networks over time. One such work  \cite{ZhouYangRenWuZhuang:2018:Dynamicnetworkembeddingbymodelingtriadicclosureprocess} uses the triadic closure property where it tries to model the probability that an open triad in the current time-step will become closed in future. Recent focus of dynamic network representation learning is on adapting graph neural network based static representation learning methods \cite{HamiltonYingLeskovec:2017:InductiveRepresentationLearningonLargeGraphs, petarEtAl:2018:Graphattentionnetworks, ZitnikEtAL:2018:Modelingpolypharmacysideeffectswithgraphconvolutionalnetworks} to the case of dynamic networks. For example, \citet{GoyalKamraHeLiu:2018:DynGEM:DeepEmbeddingMethodforDynamicGraphs} use an incremental learning of graph auto-encoder at each time-step. The above mentioned approaches can only model short term dynamics in the networks. For capturing long-term dynamics, recurrent neural network and temporal attention is applied. In
\cite{goyalChhetriCanedo:2020:dyngraph2vec:Capturingnetworkdynamicsusindynamicgraphrepresentationlearning}, graph encoders are used with recurrent neural network for modelling the evolution of node embeddings. In \cite{SankarWuGouZhangYang:2020:DySAT:DeepNeuralRepresentationLearningonDynamicGraphsviaSelf-AttentionNetworks}, graph attention networks are used with dynamic self-attention for capturing long-term dynamics of nodes.

% =============================================== %

\paragraph{Temporal KGs: }
The model proposed in this paper is an extrapolation method for temporal KGs. Here, we forecast the future links by observing the dynamics in the temporal KG in the past. In an existing approach known as Know-Evolve \cite{trivediEtAL:2017:Know-Evolve:DeepTemporalReasoningforDynamicKnowledgeGraphs}, temporal point process is used for predicting links in a temporal KG. Here, the authors assume that the properties of entities evolve as they interact with other entities, and use these entity representations as an input to a bilinear scoring function defined by a relationship matrix. The score of the bilinear function is used as a conditional density in the temporal point process. This model is extended in DyRep \cite{trivediEtAL:2019:DyRep:LearningRepresentationsoverDynamicGraphs}, where a attention based neighborhood aggregator is added to the entity evolution model. In a recent work RE-NET \cite{WoojeongHeQuChenZhangSzekelyRe:2019:RecurrentEventNetwork:GlobalStructureInferenceOverTemporalKnowledgeGraph}, a graph convolution encoder was used to aggregate the neighborhood information at each time-step in the past. A recurrent neural network then aggregates the historical neighborhood information which is then used for forecasting the probability of links in the KG in future.

% =============================================== %

\section{Conclusion}
In this paper, we presented a new statistical model called Neural Latent Space Model (NLSM) for dynamic networks. Unlike most existing approaches which focus on modelling homogeneous dynamic networks, our approach seamlessly works for both homogeneous as well as heterogeneous dynamic networks. This is achieved by using relation specific interaction matrices for modelling links. For heterogeneous networks like the temporal KGs which have multiple type of relations, our model has an interaction matrix for each relation type. We also developed a neural network based variational inference procedure for performing inference in NLSM. Through our experiments, we demonstrated the utility of our approach by using it to perform link forecasting where we achieved state-of-the-art performance on several datasets in both single relation and multi-relational setting.

\section*{Acknowledgements}
This work arose from a visit of Ambedkar Dukkipati to Eurandom, with the generous support of the STAR cluster and Eurandom. The authors TG, SG, AK and AD would like to thank the Ministry of Human Resource Development (MHRD), Government of India, for their generous funding towards this work through the UAY Project: {IISc 001}. 

%%%%%%%%%%%%%%%%%%%%%%%%%%%%%%%%%%%%%%%%%%%%%%%%%%%%%%%%%%%%%%%%%%%%%%%%%%%%%%%

%\bibliographystyle{aaai21}
\bibliography{biblio}

% =============================================== %

\newpage

\appendix

\section*{Appendix A: Figures}
\begin{figure}
    \centerline{\includegraphics[scale=1.2]{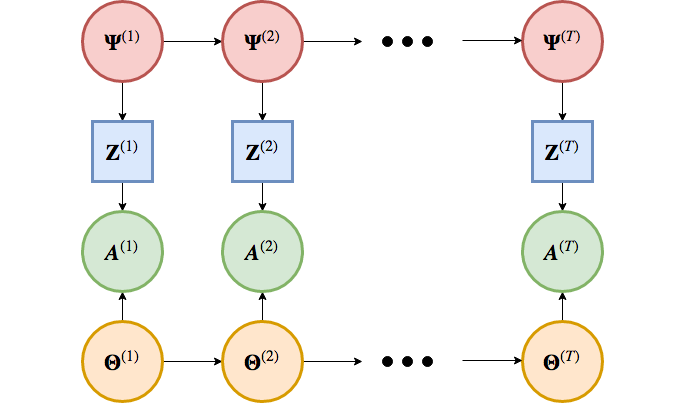}}
    \caption{Graphical model for NLSM. The vectors $\bm{\Psi}^{(t)}$ are used to model the latent node attribute vectors and $\bm{\Theta}^{(t)}$ represent the interaction matrices. Note that elements of $\mathbf{Z}^{(t)}$ are a deterministic function of the corresponding elements of $\bm{\Psi}^{(t)}$.}
    \label{fig:graphical_model}
\end{figure}

\begin{figure}
    \centerline{\includegraphics[scale=1.5]{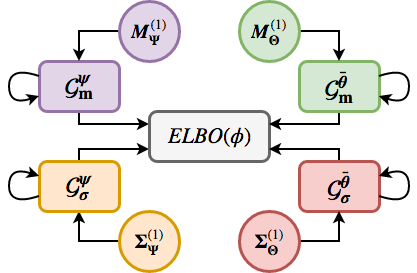}}
    \caption{Inference Network Architecture. $\mathbf{M_x}^{(1)}$ and $\bm{\Sigma_x}^{(1)}$ represent the initial learnable embeddings that are used by the respective GRUs $\bm{\mathcal{G}_m}^x$ and $\bm{\mathcal{G}_\sigma}^x$. The output of all GRUs is used to compute ELBO.}
    \label{fig:network_architecture}
\end{figure}

The graphical model for NLSM is shown in Figure \ref{fig:graphical_model}. The inference architecture is shown in Figure \ref{fig:network_architecture}.

% ============================================= %

\section*{Appendix B: ELBO derivation}
Recall that we want to approximate the posterior distribution $P(\mathbf{H} | \mathbf{X})$ using a parameterized distribution $Q_{\bm{\Phi}}(\mathbf{H})$. We quantify the quality of approximation by using the KL-divergence $KL(Q_{\bm{\Phi}}(\mathbf{H}) \vert\vert P(\mathbf{H} | \mathbf{X}))$ between $Q_{\bm{\Phi}}$ and $P$. Thus, our objective is to minimize this KL-divergence. We abbreviate $KL(Q_{\bm{\Phi}}(\mathbf{H}) \vert\vert P(\mathbf{H} | \mathbf{X}))$ by $KL(Q \vert\vert P)$ and suppress the dependence of $Q$ on $\bm{\Phi}$ in the notation to avoid clutter. Note that
\begin{align}
    KL(Q \vert \vert P) &= E_{\mathbf{H} \sim Q}\Big[\log \frac{Q(\mathbf{H})}{P(\mathbf{H}|\mathbf{X})}\Big] \notag \\
    &= E_{\mathbf{H} \sim Q}\Big[\log \frac{Q(\mathbf{H}) P(\mathbf{X})}{P(\mathbf{H}, \mathbf{X})}\Big] \notag \\
    &= \log P(\mathbf{X}) - E_{\mathbf{H} \sim Q}[\log P(\mathbf{H}, \mathbf{X}) - \log Q(\mathbf{H})] \notag \\
    &= \log P(\mathbf{X}) - \text{ELBO}(Q),
\end{align}
where, $\text{ELBO}(Q) = E_{\mathbf{H} \sim Q}[\log P(\mathbf{H}, \mathbf{X}) - \log Q(\mathbf{H})]$. As $\log P(\mathbf(X))$ is independent of the choice of $Q$, minimizing $KL(Q \vert \vert P)$ is equivalent to maximizing the ELBO.

% ============================================= %

\section*{Appendix C: Dataset Description}

\paragraph{Communication Networks: }We use \cite{Klimt2004IntroducingTE} Enron and UCI \cite{uci2009} datasets. Enron contains information about emails that were exchanged between individuals belonging to an European research organization. The links in UCI represent messages exchanged between users in an online platform. 

\paragraph{Rating Networks: }We use Yelp \footnote{https://www.yelp.com/dataset/challenge} and ML-10M \cite{HarperEtAL:2015:TheMovieLensDatasets:HistoryandContext} datasets. These are dynamic bipartite graphs comprising of two types of nodes. In Yelp, edges represent rating links between users and businesses, and in ML-10M edges appear between users and tags they apply to the movies.

\paragraph{Temporal KGs: } We used WIKI \cite{wiki2018} and YAGO \cite{Mahdisoltani2014YAGO3AK} datasets. These are heterogeneous graphs where facts are stored in quadruplets of the form \texttt{(Subject Entity, Relation, Object Entity, Time)} denoted by $(s; r; o; t )$. This means that the link \texttt{ (Subject Entity, Relation, Object Entity)} is valid at time $t$. For both WIKI and YAGO, time $t$ is represented in the granularity of years.

For Enron, UCI, and Yelp, we used the preprocessed datasets used in \cite{SankarWuGouZhangYang:2020:DySAT:DeepNeuralRepresentationLearningonDynamicGraphsviaSelf-AttentionNetworks}. For WIKI and YAGO, we used the preprocessed datasets used in \cite{WoojeongHeQuChenZhangSzekelyRe:2019:RecurrentEventNetwork:GlobalStructureInferenceOverTemporalKnowledgeGraph}. We used the same training and test splits used by the respective works. 

% ============================================= %

\section*{Appendix D: Evaluation Metrics}

\paragraph{Single Relation Link Forecasting}
For evaluating our link forecasting performance in communication and rating networks, we use the \textbf{AUC} metric. AUC stands for \textit{Area Under Curve} and this metric computes the area under the true-positive rate vs false-positive rate curve for various values of threshold used for classification. Values close to $1$ indicate good results. For aggegrating the forecasting performance across the time-steps, we use the Micro-AUC and Macro-AUC variants of the metric. Micro-AUC is calculated considering all the links across the time-steps and Macro-AUC is the average of AUC at each time-step.

\paragraph{Multi-Relational Link Prediction}
We use the well known Mean Reciprocal Rank (\textbf{MRR}), \textbf{Hits@3} and \textbf{Hits@10} metrics. For calculating these scores, in each quadruplet in Temporal KG, we find the rank of the subject entity based on the probability of link formation when replaced with other entities in the dataset. Here, ranking is from 1 to N with the rank decreasing as the probability of the entity becomes higher. Similarly, we find the rank of object entity as well. These ranks are then used for calculating \textbf{MRR} and \textbf{Hits@k}. \textbf{MRR} is calculated by taking mean reciprocal of ranks and \textbf{Hits@k} is calculated by finding the percentage of times true entity rank is not more than $k$. Further, when we replace the subject/object entity with other entities in the dataset, there is a chance that the new quadruplet is a valid link in the dataset. This might lead to entities getting higher ranks because there are other valid entities which have higher probability. So, to consider this while calculating the metric, we calculate the filtered version of earlier metrics. Here, when we replace the subject/object entity with other entities, we ignore entities which lead to a valid link in the dataset. These metric scores are denoted by \textbf{MRR-filtered}, \textbf{Hits@3-filtered} and \textbf{Hits@10-filtered}, while earlier metrics scores are denoted by \textbf{MRR-raw}, \textbf{Hits@3-raw} and \textbf{Hits@10-raw}. 

% =============================================== %

\section*{Appendix E: Details about Experimental Setup}

For models NLSM and BAS, we use the ADAM \cite{KingmaBa:2014:AdamAMethodforStochasticOptimization} optimizer in Pytorch \cite{paszke:2017:AutomaticdifferentiationinPyTorch} and chose a learning rate of $0.01$ for all experiments. We fix the hyper-parameters $s_\theta = 0.1$, $s_\psi = 0.1$, $\sigma_\theta= 10$ and $\sigma_\psi =10$ in all the experiments. For learning in communication and rating networks, we use a batch size $b=20$ edges per time-step and for each positive edge we sample a negative edge from the dataset. The training is done for $1000$ epochs for each of the time steps. In Temporal Knowledge Graphs (KG), we use the same batch size as before and for each edge, which is now a quadruplet, we use two negative samples, one for Subject Entity and the other one for Object Entity. We then train for $2000$ epochs at each of the time steps.

\paragraph{Computational Infrastructure. } We used a workstation with Intel Xeon processor and a Nvidia-DGX-1 with 251 GB of main memory and 4 x Tesla V100 GPUs each having 32 GB memory.

% =============================================== %

\begin{table*}[t]
\begin{center}
\begin{tabular}{lccccc}
    \toprule
    & \textbf{LFRM}$^a$ & \textbf{DRIFT}$^b$ & \textbf{DMMG}$^c$ & \textbf{iELSM}$^d$ & \textbf{NLSM} (ours) \\
    \midrule
    \textbf{Enron-A} & 0.777 & 0.910 & - & 0.913 & \textbf{0.923 $\pm$ 0.002} \\
    \textbf{Enron-B} & - & - & - & - & \textbf{0.928 $\pm$ 0.004} \\
    \textbf{Infocom} & 0.640 & 0.782 & 0.804 & \textbf{0.868} & 0821 $\pm$ 0.007 \\
    \textbf{NIPS-110} & 0.398 & 0.672 & 0.732 & 0.754 & \textbf{0.810 $\pm$ 0.008} \\
    \textbf{EU-U} & - & - & - & 0.948 & \textbf{0.973 $\pm$ 0.001} \\
    \textbf{EU-D} & - & - & - & - & \textbf{0.934 $\pm$ 0.003} \\
    \textbf{CollegeMsg} & - & - & - & - & \textbf{0.857 $\pm$ 0.007} \\
    \bottomrule
\end{tabular}
\end{center}

\caption{Additional single-step link forecasting experiments: Mean AUC scores with standard deviation across 20 independent executions of the experiment. $*$ indicates that the network is directed. \textbf{Citations: }$^a$\citep{MillerEtAl:2009:NonparametricLatentFeatureModelsForLinkPrediction}, $^b$\citep{FouldsEtAl:2011:ADynamicRelationalInfiniteFeatureModelForLongitudinalSocialNetworks}, $^c$\citep{KimEtAl:2013:NonparametricMultiGroupMembershipModelForDynamicNetworks}, $^d$\citep{GuptaEtAl:2018:EvolvingLatentSpaceModelForDynamicNetworks}}
\label{table:smaller_datasets}

\end{table*}

\section*{Appendix F: Additional Experiments}

\subsection*{Results on Smaller Datasets}

We also compare the performance of our model on smaller datasets against more traditional approaches which work especially well for these datasets. For these experiments, we use the following datasets:

\textbf{1. Enron email: }The full Enron email corpus \citep{KlimtEtAl:2004:TheEnronCorpus} has 149 nodes corresponding to employees in a company. A directed edge from node $i$ to node $j$ implies that $i$ sent an email to $j$. We use an undirected subset of the Enron corpus (Enron-A) consisting of 50 nodes as described in \citep{FouldsEtAl:2011:ADynamicRelationalInfiniteFeatureModelForLongitudinalSocialNetworks}. We also perform link prediction on the full, directed Enron corpus with $149$ nodes by taking data from years 2000-2001 where each network snapshot corresponds to a month (Enron-B). 

\textbf{2. Infocom: }There are 78 nodes in this network. An undirected edge from node $i$ to node $j$ at timestep $t$ indicates that $i$ and $j$ were in proximity of each other during that timestep. We obtain a dynamic network with 50 snapshots by using the procedure outlined in \citep{GuptaEtAl:2018:EvolvingLatentSpaceModelForDynamicNetworks}.

\textbf{3. NIPS co-authorship: }This dataset consists of 5,722 nodes. An undirected edge from node $i$ to node $j$ indicates that $i$ and $j$ co-authored a paper. We consider a subset of the dataset containing 110 nodes as described in \citep{HeaukulaniEtAl:2013:DynamicProbabilisticModelsForLatentFeaturePropagationInSocialNetworks}. We refer to this dataset as NIPS-110.

\textbf{4. EU Email: }This dataset \citep{YinEtAl:2017:LocalHigherOrderGraphClustering} contains information about emails that were exchanged between individuals belonging to an European research organization. There are $986$ nodes in the network. We consider the first $495$ days and create $33$ network snapshots by aggregating data over $15$ day time windows. We treat this as an undirected (EU-U) as well as a directed (EU-D) network.

\textbf{5. CollegeMsg: }There are 1899 nodes in this dataset \citep{PanzarasaEtAl:2009:PatternsAndDynamicsOfUsersBehaviourAndInteraction}. A binary, directed edge corresponds to a message exchanged between the sender and receiver. Temporal data for 193 days is available. We discard the last 3 days and divide the data into 10 days wide buckets which gives us 19 snapshots.

We set $K=32$, $s_\theta = s_\psi = 0.1$ and $\sigma_\theta = \sigma_\psi = 10$ and consider single-step link forecasting. For comparison, we use  \citep{MillerEtAl:2009:NonparametricLatentFeatureModelsForLinkPrediction,FouldsEtAl:2011:ADynamicRelationalInfiniteFeatureModelForLongitudinalSocialNetworks,KimEtAl:2013:NonparametricMultiGroupMembershipModelForDynamicNetworks,GuptaEtAl:2018:EvolvingLatentSpaceModelForDynamicNetworks}. LFRM or Latent Feature Infinite Relational Model \citep{MillerEtAl:2009:NonparametricLatentFeatureModelsForLinkPrediction} represents nodes in a static network using binary feature vectors. It is a non-parametric model. It imposes Indian Buffet Process \citep{Griffiths:2011:TheIndianBuffetProcessAnIntroductionAndReview} prior on a feature matrix that encodes feature vector of nodes in its rows. \citep{FouldsEtAl:2011:ADynamicRelationalInfiniteFeatureModelForLongitudinalSocialNetworks} proposed Dynamic Relational Infinite Feature Model (DRIFT) as an extension of LFRM for dynamic networks by allowing features of nodes to evolve under Markov assumption. While computing predictions for time $t$, LFRM model trained on time $t-1$ was used \citep{FouldsEtAl:2011:ADynamicRelationalInfiniteFeatureModelForLongitudinalSocialNetworks}. Dynamic Multigroup Membership Model (DMMG) \citep{KimEtAl:2013:NonparametricMultiGroupMembershipModelForDynamicNetworks} uses a model similar to ours but with discrete node attributes and fixed interaction matrices. All these methods use MCMC based inferece but since our model is completely differentiable, we are able to use a neural network based variational inference procedure. \citep{GuptaEtAl:2018:EvolvingLatentSpaceModelForDynamicNetworks} also have a differentiable model for which they use neural network based variational inference, however, their generative model is restrictive as they only focus on assortative and undirected networks.

Table \ref{table:smaller_datasets} reports the  AUC scores obtained by first averaging the snapshot-wise scores and then taking the mean values across 20 independent runs of the inference network. Missing entries are either because of the incompatibility of the method with the network type (for instance, the method may not support a directed network) or because of an unavailability of the implementation. It can be seen that our approach outperforms all the other approaches on all datasets except Infocom. We believe that this is because the Infocom network changes quickly across snapshots as it is a contact network and it has abrupt breaks (at the end of each day when participants leave the premises). This violates our assumption of a slowly changing network.

% =============================================== %

\subsection*{Changing Latent Vector Dimension $K$}

\begin{figure}[t]
    \centering
    \includegraphics{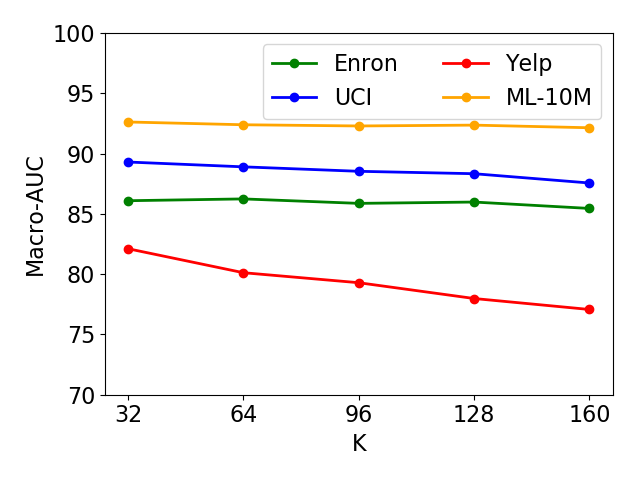}
    \caption{[Best viewed in color] Single link forecasting performance in Macro-AUC with different values of latent vector size $K$. (Macro AUC in \% averaged over 10 runs)  }
    \label{fig:albationKmacroAUC}
\end{figure}

\begin{figure}[t]
    \centering
    \includegraphics{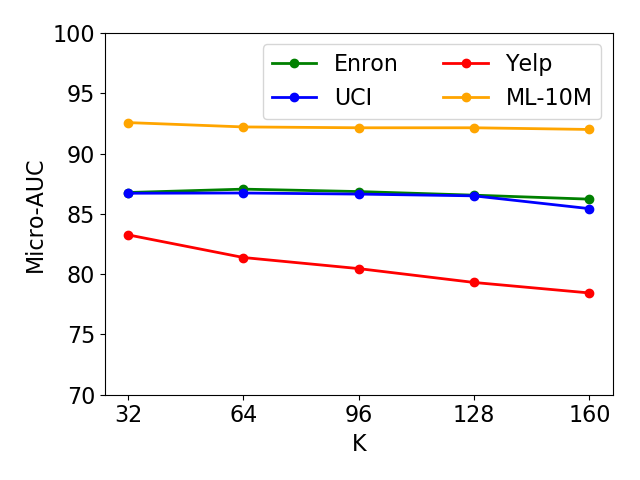}
    \caption{[Best viewed in color] Single link forecasting performance in Micro-AUC with different values of latent vector size $K$ }
    \label{fig:albationKmicroAUC}
\end{figure}

We varied the latent vector dimension $K$ to study its impact on the link forecasting performance. From Figure \ref{fig:albationKmacroAUC}, we can observe that smaller values of $K$ give better performance. This is because larger values of $K$ cause over-fitting for these datasets. We can also observe similar trends in Micro-AUC scores in Figure \ref{fig:albationKmacroAUC}.

% =============================================== %

\section*{Appendix G: Qualitative Analysis}

\begin{figure*}[t]
    \fbox{\subfigure[$t=1$]{\label{fig:nips_case_study_t1}\includegraphics[width=0.24\linewidth]{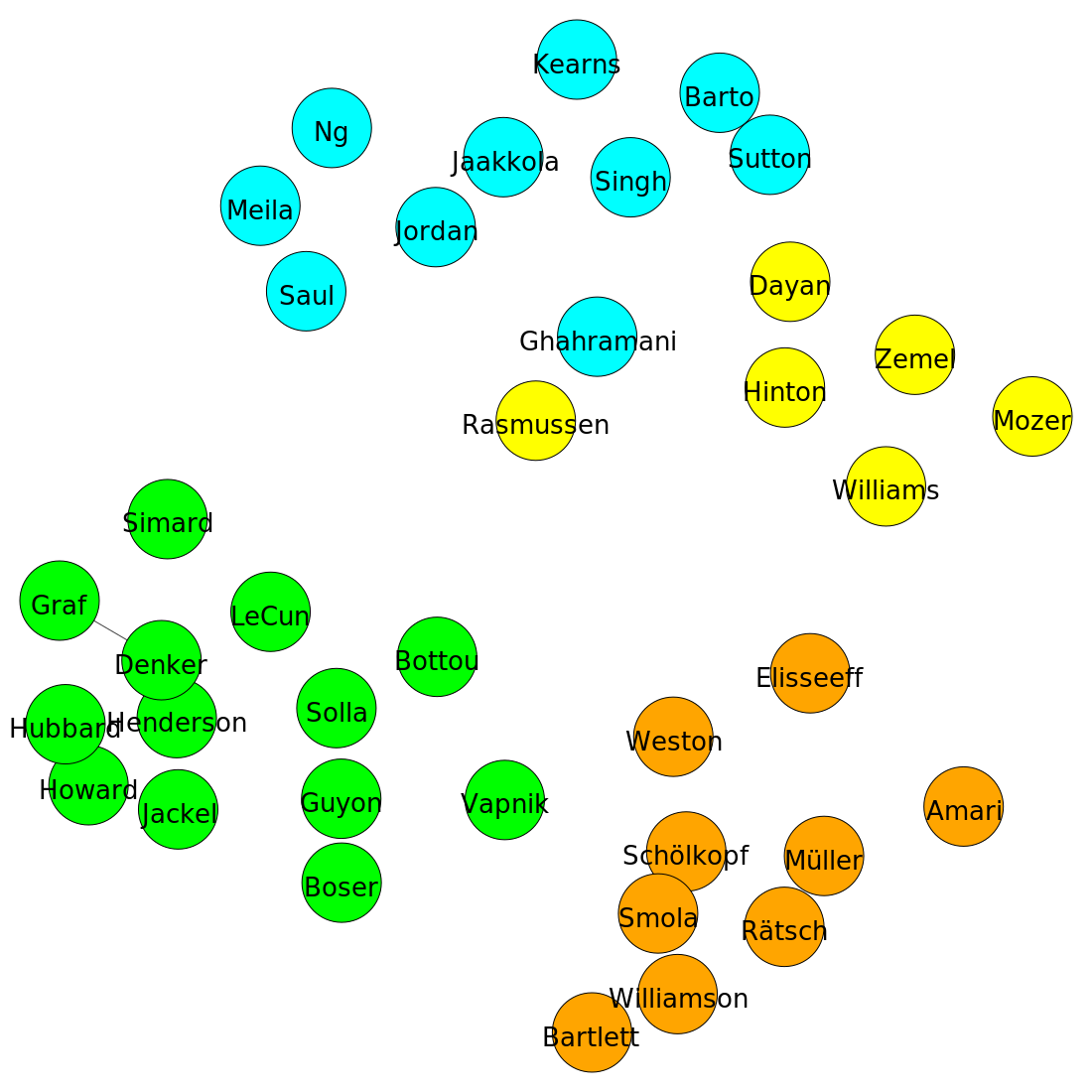}}}%
    \fbox{\subfigure[$t=3$]{\label{fig:nips_case_study_t3}\includegraphics[width=0.24\linewidth]{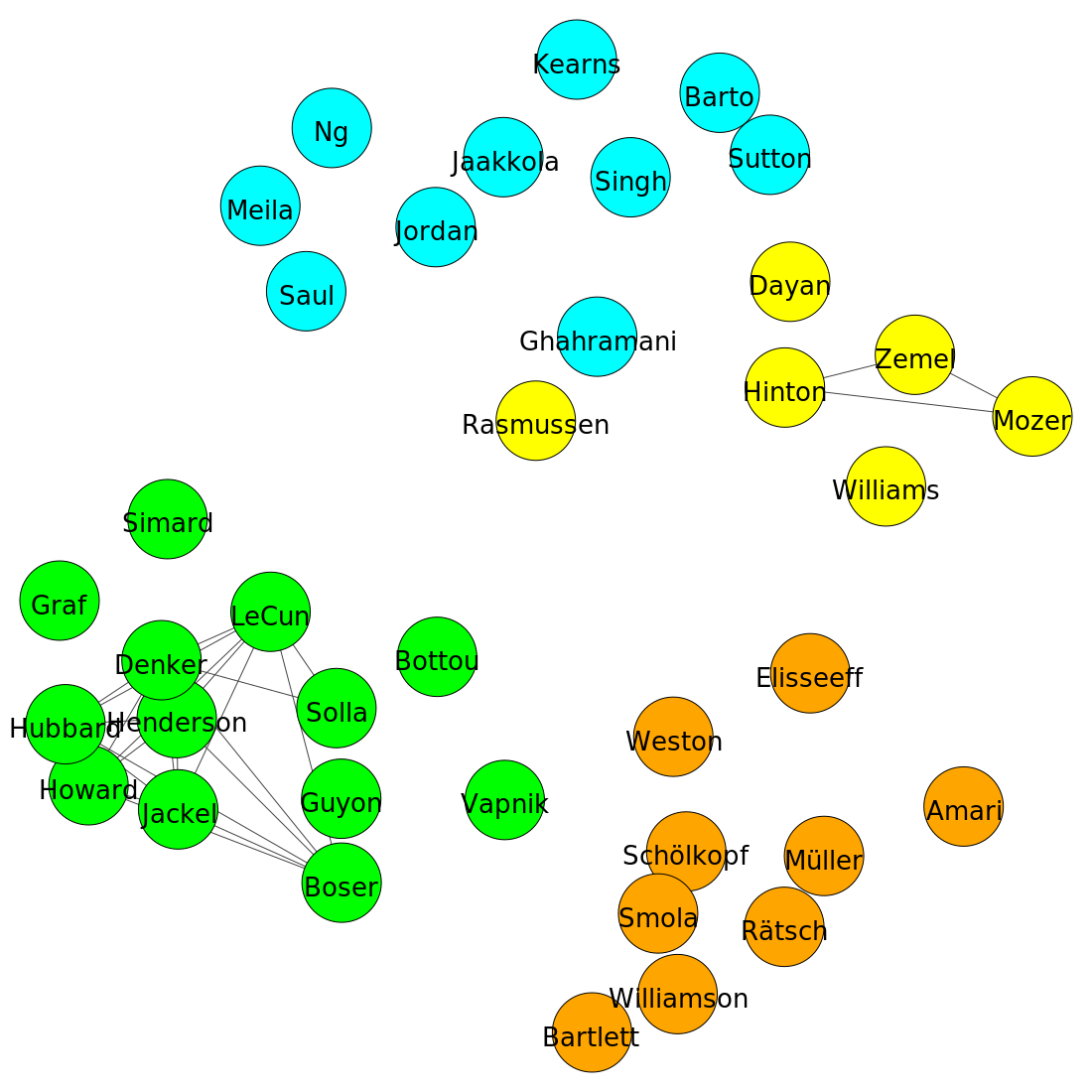}}}%
    \fbox{\subfigure[$t=7$]{\label{fig:nips_case_study_t7}\includegraphics[width=0.24\linewidth]{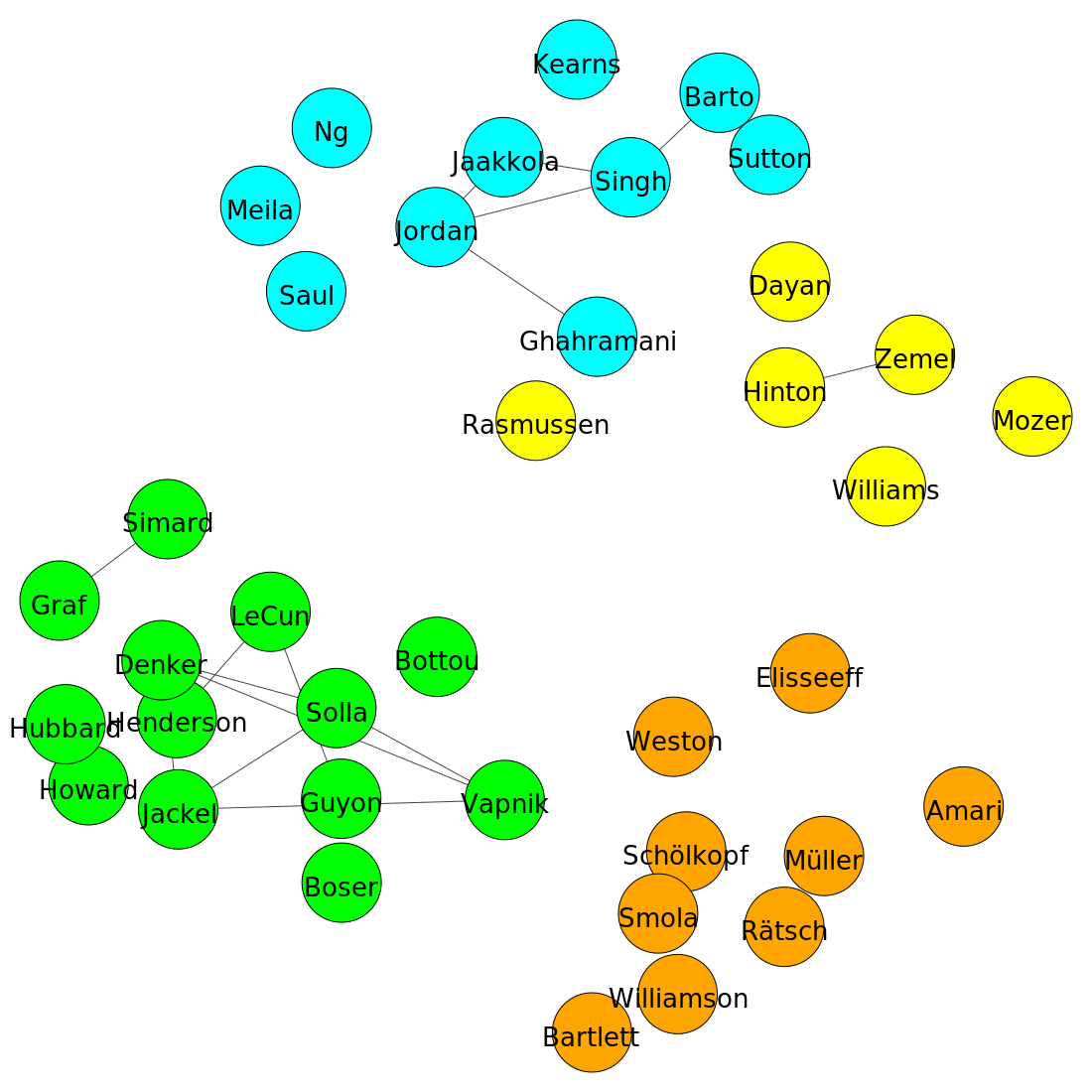}}}%
    \fbox{\subfigure[$t=11$]{\label{fig:nips_case_study_t11}\includegraphics[width=0.24\linewidth]{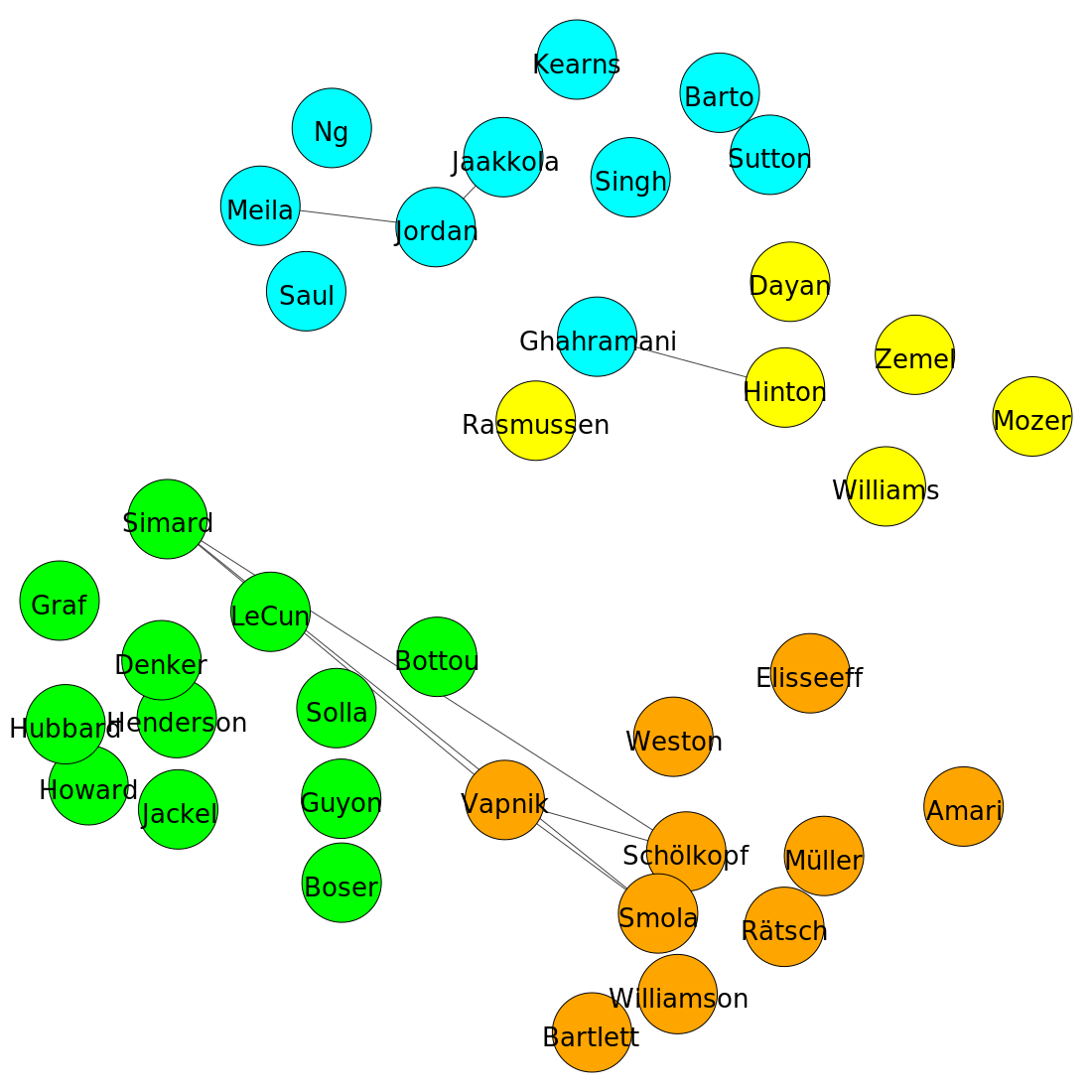}}}%
    \caption{[Best viewed in color] Network between a subset of authors chosen to highlight the community structure. Different colors have been used to differentiate the communities found by running spectral clustering algorithm on a normalized version of adjacency matrix predicted by our method.}\label{fig:nips_case_study}
\end{figure*}

\begin{figure*}[!ht]
    \fbox{\subfigure{\includegraphics[width=0.24\linewidth]{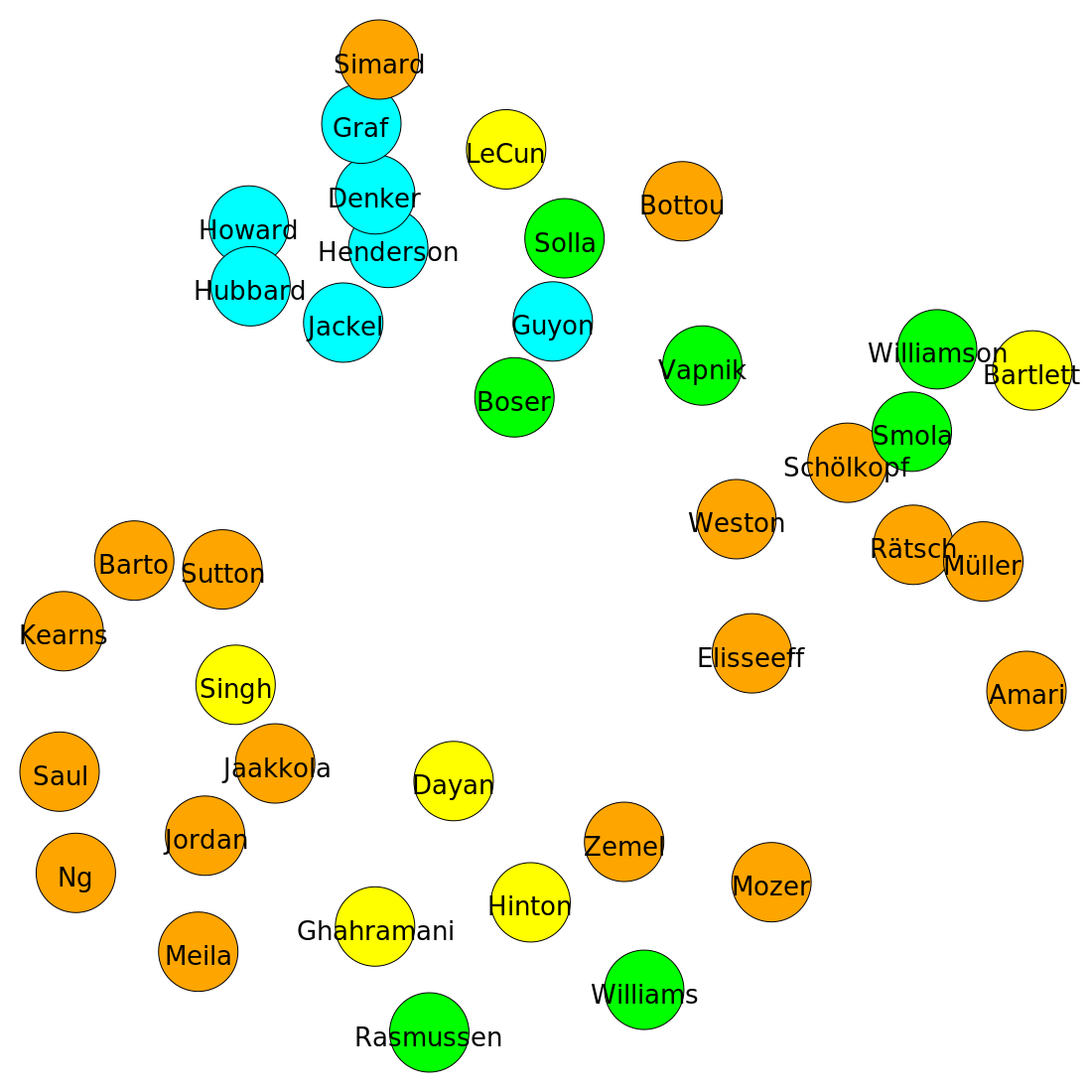}}}%
    \fbox{\subfigure{\includegraphics[width=0.24\linewidth]{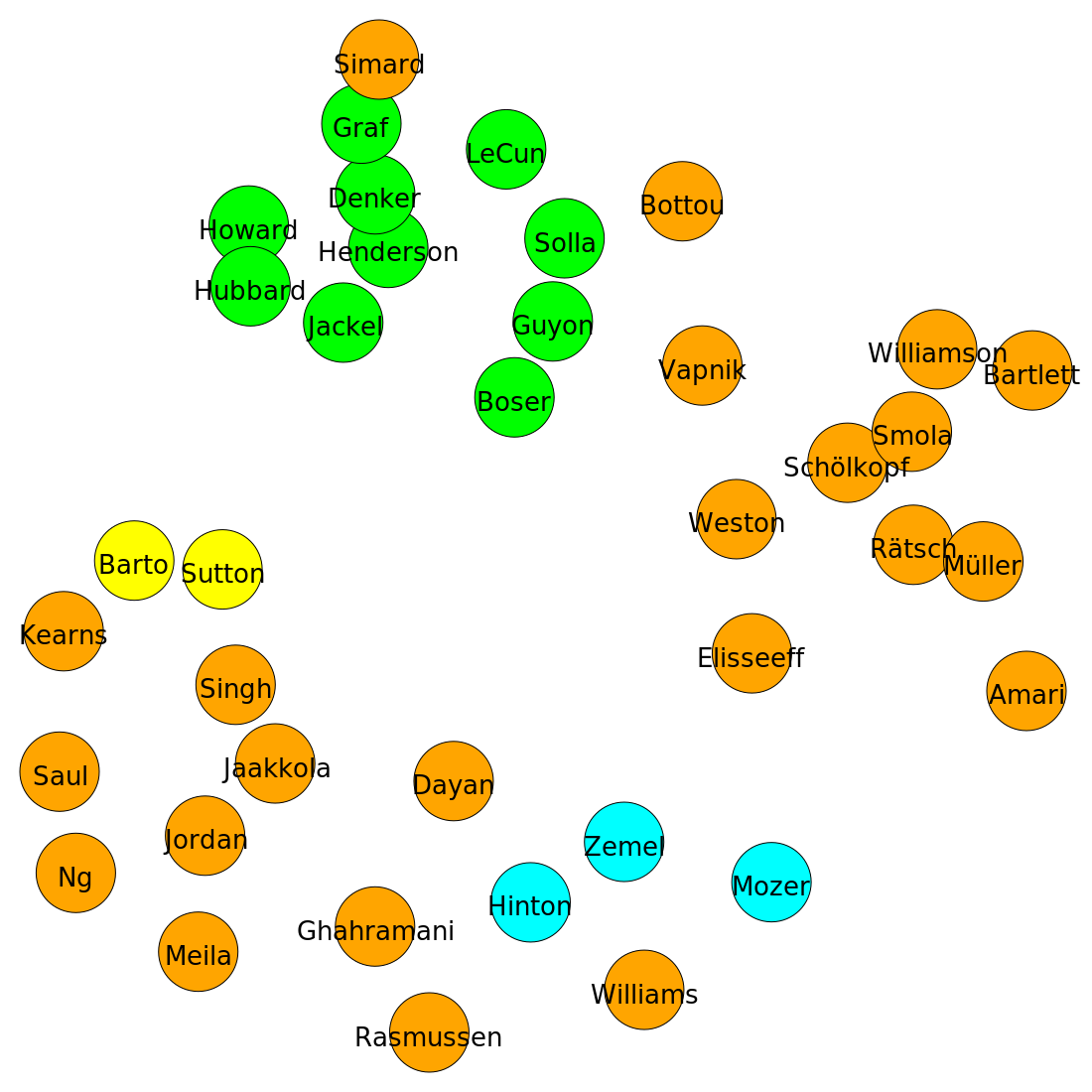}}}%
    \fbox{\subfigure{\includegraphics[width=0.24\linewidth]{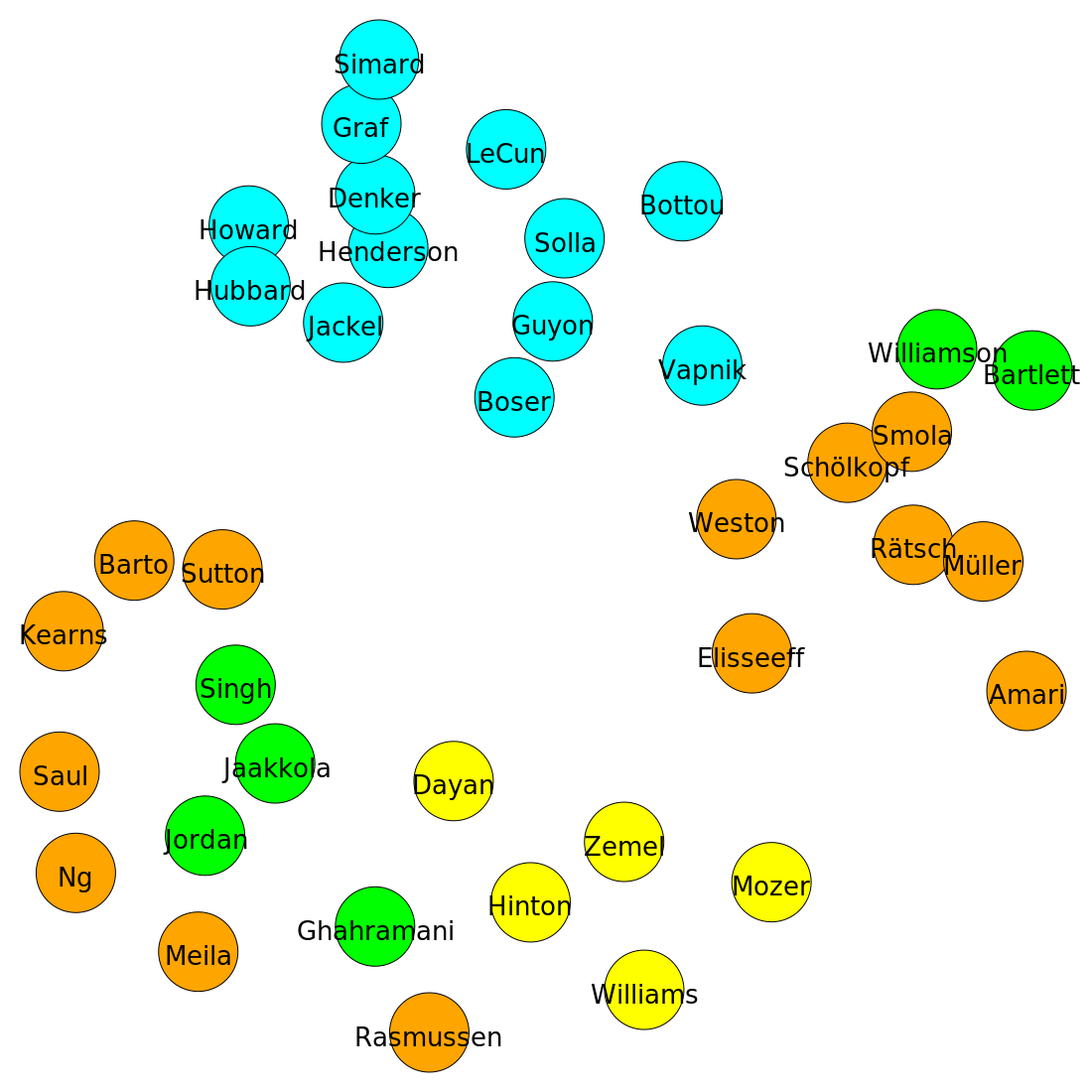}}}%
    \fbox{\subfigure{\includegraphics[width=0.24\linewidth]{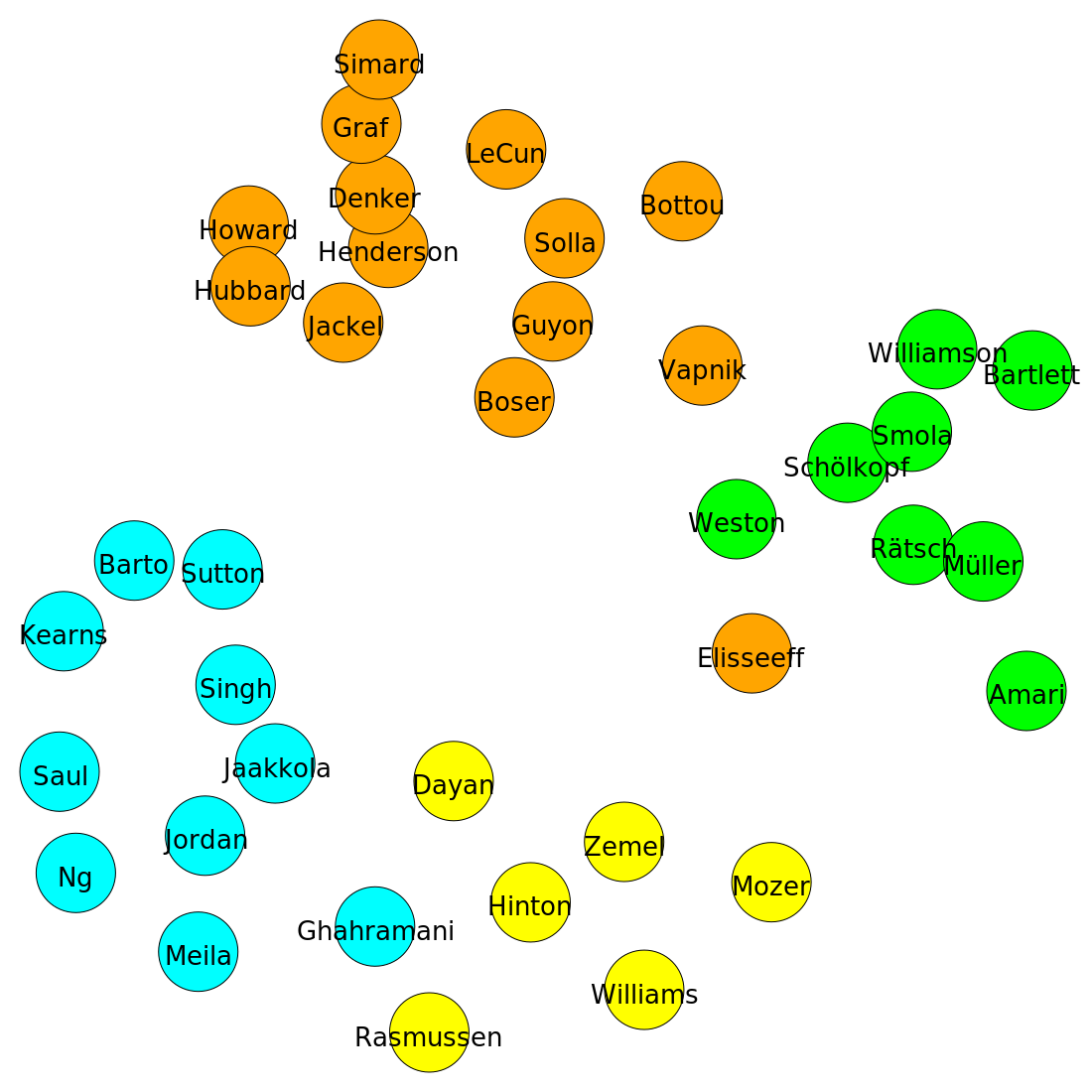}}}
    \fbox{\subfigure[$t=2$]{\includegraphics[width=0.24\linewidth]{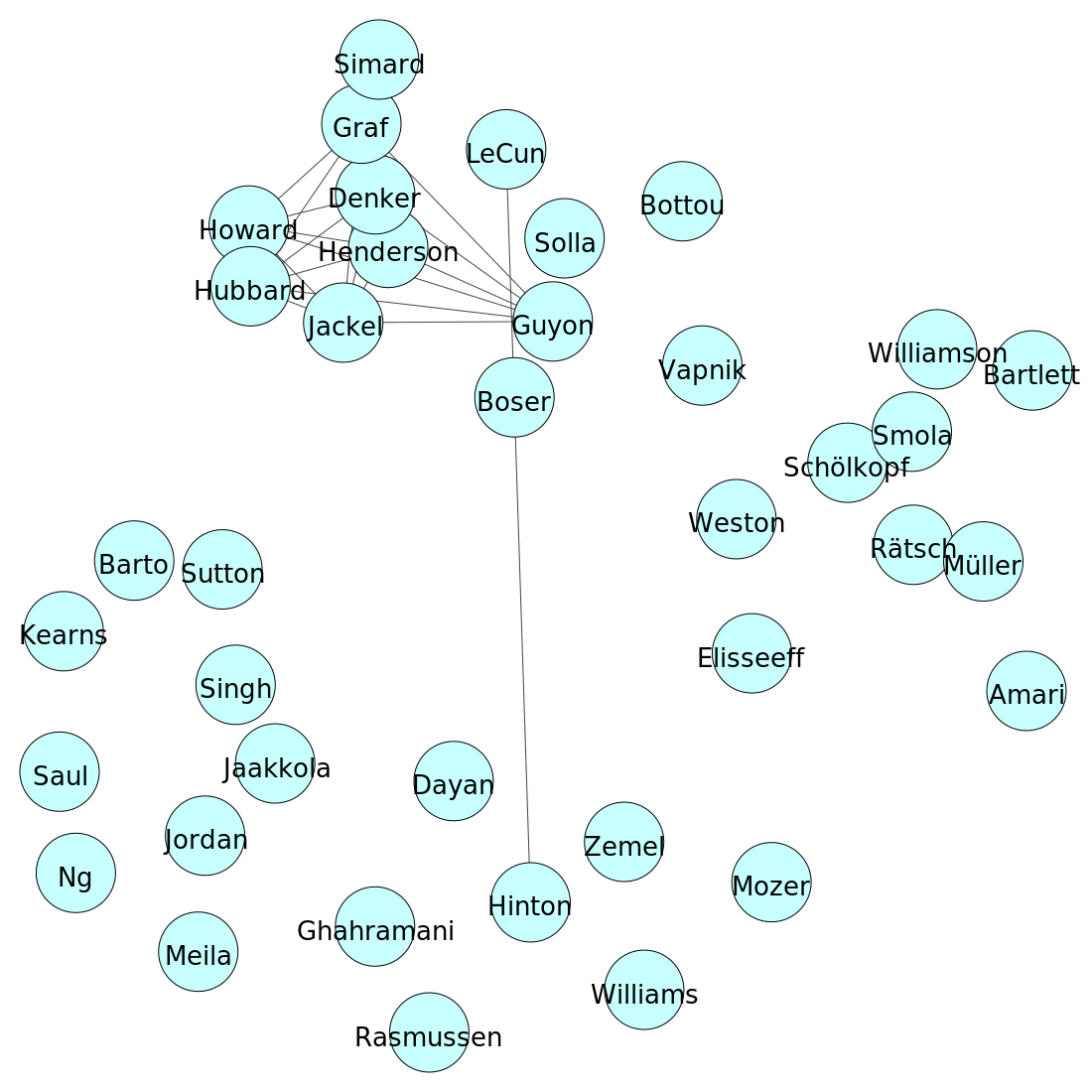}}}%
    \fbox{\subfigure[$t=3$]{\includegraphics[width=0.24\linewidth]{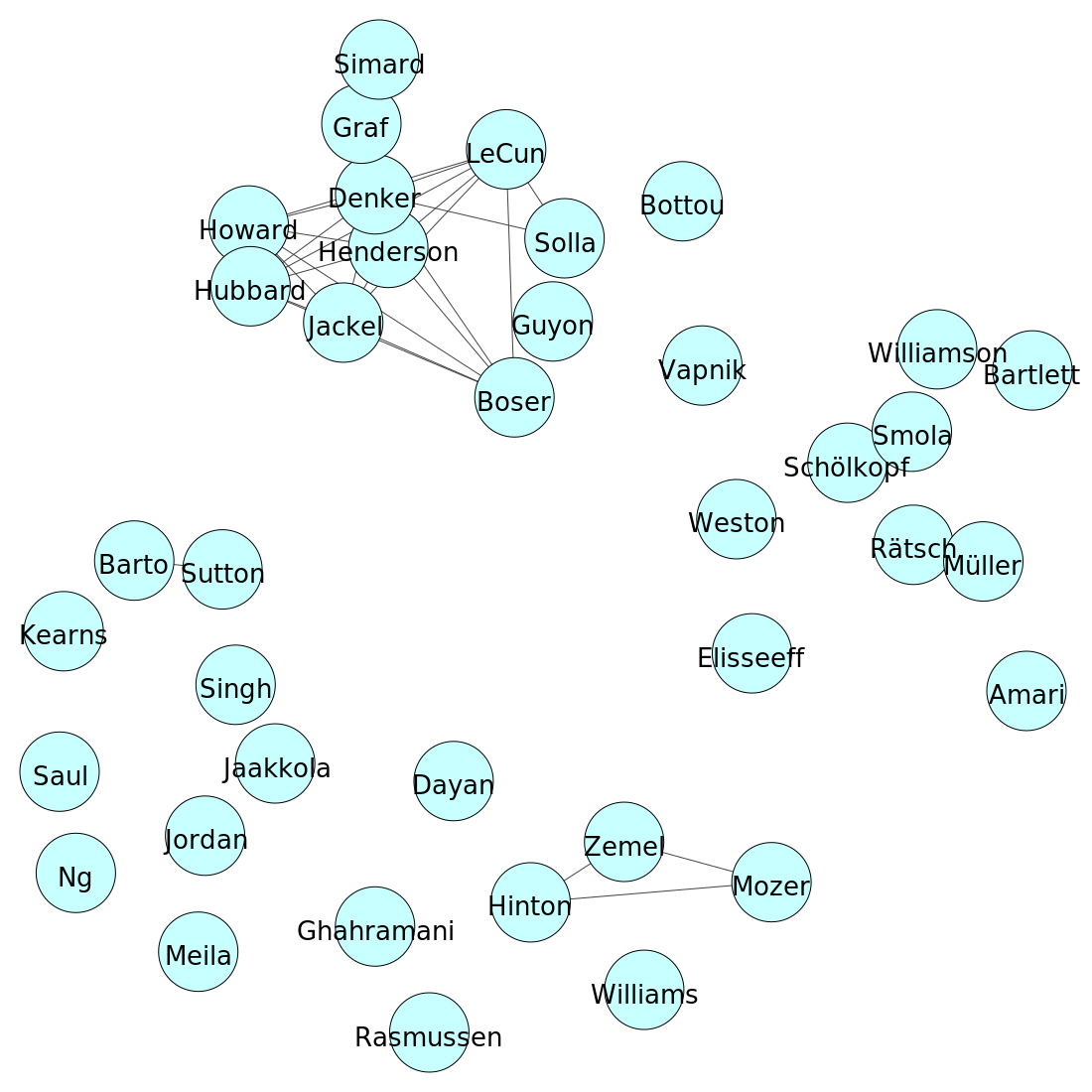}}}%
    \fbox{\subfigure[$t=7$]{\includegraphics[width=0.24\linewidth]{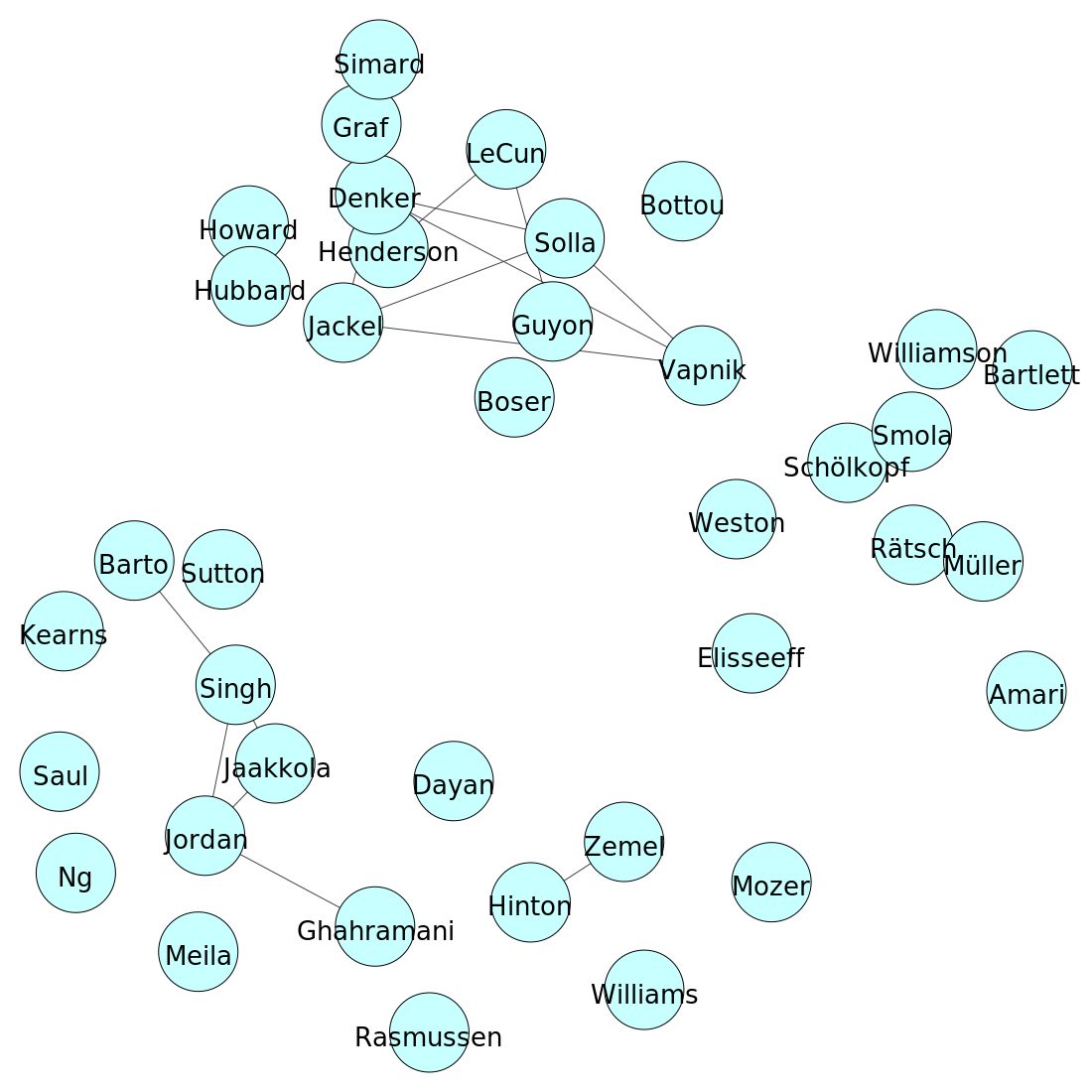}}}%
    \fbox{\subfigure[$t=13$]{\includegraphics[width=0.24\linewidth]{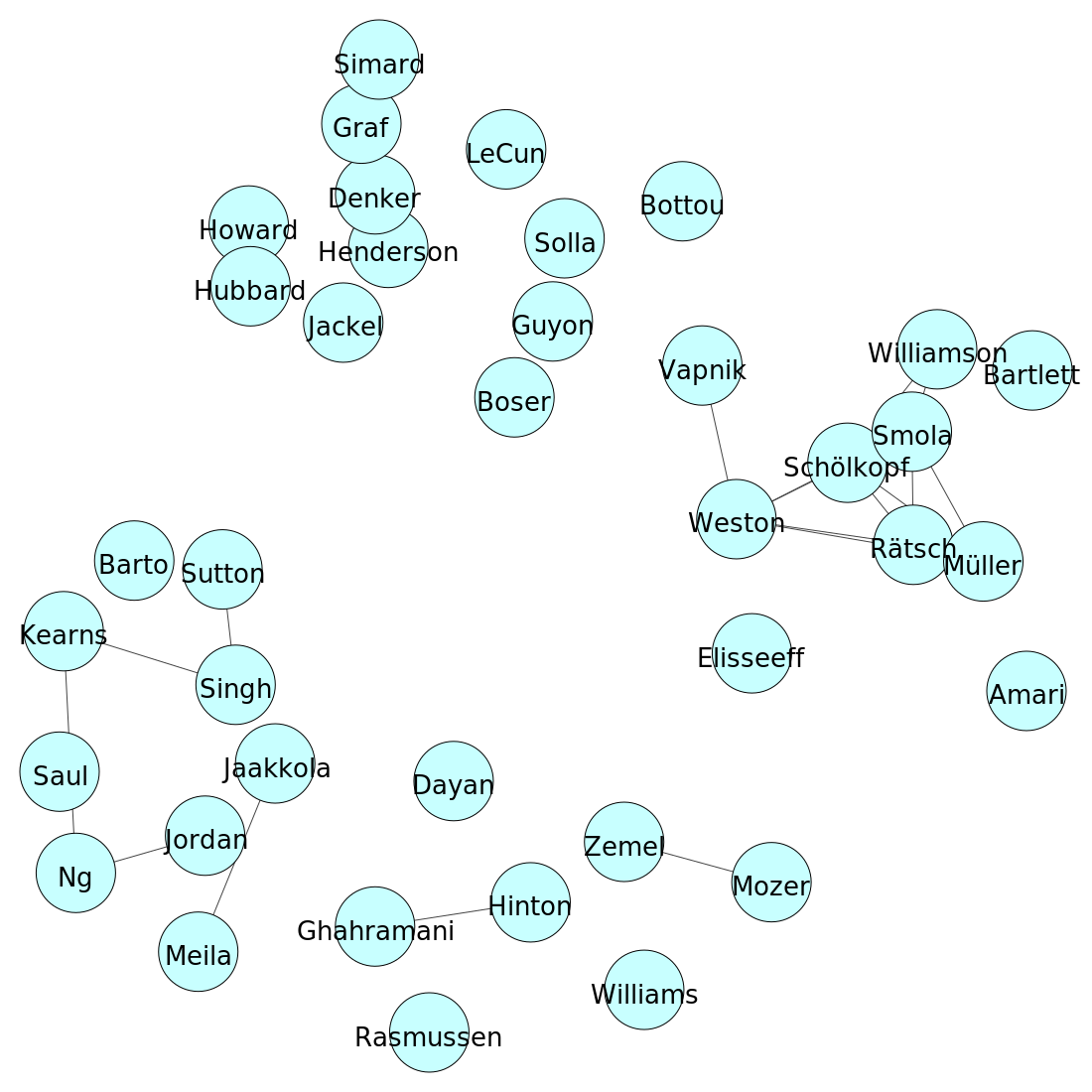}}}%
    \caption{[Best viewed in color] First row depicts the outcome of community detection on the first time step. Second row shows the last adjacency matrix observed by the corresponding inference network that was used to provide embeddings for performing community detection in first row.}\label{fig:nips_case_study_2}
\end{figure*} 

In this section, we present some qualitative insights about NIPS-110 and MIT Reality Mining datasets that were revealed by our model. For NIPS-110, author names were obtained by parsing the raw data\footnote{\url{http://www.cs.huji.ac.il/~papushado/nips_collab_data.html}} and selecting top 110 authors as in \cite{KimEtAl:2013:NonparametricMultiGroupMembershipModelForDynamicNetworks}. We use $K = 8$ for this analysis. A smaller value of $K$ was chosen to aid the manual inspection process.

As ground truth communities are available for MIT Reality Mining dataset, we used it for a sanity check. It is known that two communities that align with ground truth communities can be discovered from the network structure \cite{Xu:2014:DynamicStochasticBlockmodelsForTimeEvolvingSocialNetworks,EagleEtAl:2006:TowardsTimeAwareLinkPredictionInEvolvingSocialNetworks}. We followed the same procedure as in \cite{Xu:2014:DynamicStochasticBlockmodelsForTimeEvolvingSocialNetworks} and our model was able to recover both communities. We also observed that both node latent vectors and interaction matrices evolved with time.

For NIPS-110 dataset, we observed that node latent vectors for authors did not change noticeably over time, however, the interaction matrices showed time dependent behavior. This aligns with what one might intuitively expect: authors typically do not dramatically change their domain of expertise over time, but they may start collaborating with different people as connections among different fields emerge.

We further conducted two experiments. First, we trained our inference network on all available snapshots and performed community detection on all snapshots using the trained embeddings. Second, we incrementally trained $T - 2$ inference networks (starting by observing only two snapshots for the first network and going up to observe $T-1$ snapshots for the last network), and then performed community detection on the first snapshot using embeddings for first snapshot obtained from each of the $T - 2$ trained networks.

To perform community detection at a given timestep $t$, we use the learned embeddings to compute the summation term inside $\sigma(.)$ in 
\begin{equation}
    P(a_{ij}^{(t)} = 1 | \mathbf{z}_i^{(t)}, \mathbf{z}_j^{(t)}, \{\bm{\Theta}_k^{(t)}\}_{k=1}^K) = \sigma\big(\sum_{k=1}^{K} \tilde{\theta}_k^{(t)}(i, j)\big),
\end{equation}
for all pair of nodes to get $\tilde{\mathbf{A}}^{(t)}$. We mean normalize entries of $\tilde{\mathbf{A}}^{(t)}$, exponentiate them and then perform spectral clustering on this matrix. We chose spectral clustering as it can possibly discover non-convex clusters. Note that this is different from clustering on all snapshots independently since the embeddings capture temporal smoothness.

Through the first experiment, we wish to demonstrate that learned embeddings enforce smoothness over model dynamics as evident from Fig~\ref{fig:nips_case_study}. In Fig~\ref{fig:nips_case_study_t1}, nodes have been classified into communities because they will coauthor a paper together in future, despite having no edges between them at $t=1$ (see Fig~\ref{fig:nips_case_study_t3} and \ref{fig:nips_case_study_t7}).

It might appear that nodes do not switch communities at all and that same result would have been obtained by running spectral clustering on the sum of all snapshots. However, this is not true. One can see that Vapnik is part of green community at $t=7$ and orange community at $t=11$ since after that time he publishes multiple papers with members from orange community. This demonstrates that our method captures temporal smoothness while being flexible enough to captures temporal changes.

Through the second experiment, we wish to demonstrate how learned embeddings from past are updated as new information arrives. It can be seen in Fig~\ref{fig:nips_case_study_2} that as new edges are observed in future, embeddings for first timestep are updated to reflect this information. As an example, Hinton, Williams, Zemel and Rasmussen belong to different communities when only the first two snapshots have been observed, but over time these authors become part of the same community as they publish papers together. Note that the first row in Fig~\ref{fig:nips_case_study_2} corresponds to $t=1$ for all columns, hence, new information has to temporally flow backward for Fig~\ref{fig:nips_case_study_2} to emerge.

\end{document}